\documentclass[useAMS,usenatbib]{mn2e} 
\usepackage{graphicx}

\title[Environmental Dependence of Halo Merger Rates]{Environmental Dependence of Dark Matter Halo Growth I: Halo Merger Rates}

\author[O Fakhouri and C-P Ma]{Onsi Fakhouri$^{1}$\thanks{E-mail: onsi@berkeley.edu, cpma@berkeley.edu} and Chung-Pei Ma$^{1}$\\
$^{1}$Department of Astronomy, 601 Campbell Hall, University of California, Berkeley, CA 94720}

\pagerange{ 
\pageref{firstpage}-- 
\pageref{lastpage}}

\usepackage{multirow}

\def \ZA {0} 
\def \ZB {0.51} 
\def \ZC {1.08} 
\def \ZD {2.07} 
\def \dz {\Delta z} 
\def \RATIO {\Xi}

\newcommand{\Rz}{dN_{\rm merge}/dz} 
\newcommand{\Rzf}{\frac{dN_{\rm merge}}{dz}} 
\newcommand{\ds}{\delta_7} 
\newcommand{\dst}{\delta_{7-2}} 
\newcommand{\dsfof}{\delta_{7-{\rm FOF}}} 
\newcommand{\dRfof}{\delta_{R-{\rm FOF}}} 
\newcommand{\MMIN}{1.2\times10^{12}} 
\newcommand{\ximin}{\xi_{\rm min}}

\usepackage{amsmath} 
\usepackage{multirow}

\begin{document}

\label{firstpage}

\maketitle 
\begin{abstract}
	In an earlier paper we quantified the mean merger rate of dark matter haloes as a function of redshift $z$, descendant halo mass $M_0$, and progenitor halo mass ratio $\xi$ using the Millennium simulation of the $\Lambda$CDM cosmology. Here we broaden that study and investigate the dependence of the merger rate of haloes on their surrounding environment. A number of local mass overdensity variables, both including and excluding the halo mass itself, are tested as measures of a halo's environment. The simple functional dependence on $z$, $M_0$, and $\xi$ of the merger rate found in our earlier work is largely preserved in different environments, but we find that the overall amplitude of the merger rate has a strong positive correlation with the environmental densities. For galaxy-mass haloes, we find mergers to occur $\sim 2.5$ times more frequently in the densest regions than in voids at both $z=0$ and higher redshifts. Higher-mass haloes show similar trends. We present a fitting form for this environmental dependence that is a function of both mass and local density and is valid out to $z=2$. The amplitude of the progenitor (or conditional) mass function shows a similar correlation with local overdensity, suggesting that the extended Press-Schechter model for halo growth needs to be modified to incorporate environmental effects. 
\end{abstract}

\section{Introduction} \label{introduction}

In studies of cosmological structure formation, the mass of a dark matter halo is a key variable upon which many properties of galaxies and their host haloes depend. For instance, dark matter haloes of lower mass are expected to form earlier on average than more massive haloes in hierarchical cosmological models such as $\Lambda$CDM. In semi-analytical modelling of galaxy formation (see \citealt{BaughReview} for a review), properties such as the formation redshift, halo occupation number, galaxy colour and morphology, and stellar vs AGN feedback processes are all assumed to depend on the mass of the halo (sometimes better characterised by the halo circular velocity).

In addition to the halo mass, however, recent work based on numerical simulations has shown that a halo's local environment also affects various aspects of halo formation. For instance, at a {\it fixed} mass, older haloes are found to cluster more strongly than more recently formed haloes \citep{Gottlober01, ShethTormen04, Gao2005, Harker06, Wechsler06, JingSutoMo07, WangMoJing07, GaoWhite07, Maulbetsch07}. Other halo properties such as concentration, spin, shape, and substructure mass fraction have also been found to vary with halo environment (e.g., \citealt{Avila05, Wechsler06, JingSutoMo07, GaoWhite07, Bett07}).

In contrast, no such environmental dependence is predicted in the extended Press-Schechter (EPS) and excursion set models \citep{PS74, BondEPS, LC93} that are widely used for making theoretical predictions of galaxy statistics and for Monte Carlo constructions of merger trees. The lack of environmental correlation arises from the Markovian nature of the random walks in the excursion set model. This limitation is not built into the model per se, but is an assumption stemming from the use of a tophat Fourier-space window function. When a Gaussian window function is used, for instance, \citet{ZentnerEPS} finds an environmental dependence in the halo formation redshift, but the dependence is {\it opposite} to that seen in the numerical simulations cited above. Other attempts at incorporating environmental effects into the excursion set model thus far have not been able to reproduce the correlations in simulations (e.g., \citealt{Sandvik07, DesJacques07}).

In this paper, we focus on the environmental dependence of the merger rate of dark matter haloes, a topic that has not been studied in detail. The merger rate is an important quantity for understanding and interpreting observational data on galaxy formation, growth, and feedback processes. While the mergers of galaxies and the mergers of dark matter haloes are not identical processes, the two processes are closely related, and quantifying the latter is the first key step in understanding the former. There have been few theoretical studies of merger rates (e.g., \citealt{Gottlober01, FM08, Stewart08}) probably because mergers are two- (or more-) body processes, and a large ensemble of descendent haloes {\it and} their progenitor haloes must be identified from merger trees before the rate can be reliably calculated. In comparison, studies of halo properties such as the mass function, density and velocity profiles, concentration, triaxiality, spin, and substructure distribution require only the particle information from a single simulation output.

This paper is an extension of our earlier study \cite{FM08} (henceforth FM08). There we quantified the global mean merger rates of haloes in the Millennium simulation \citep{Springel05} over a wide range of descendant halo mass ($10^{12} \la M_0 \la 10^{15} M_\odot$), progenitor mass ratio ($10^{-3} \la \xi \le 1$), and redshift ($0 \le z \la 6$). We found that when expressed in units of the mean number of mergers {\it per halo} per unit redshift, the merger rate has a very simple dependence on $M_0$, $\xi$, and $z$: the rate depends very weakly on halo mass ($\propto M_0^{0.08}$) and redshift, and scales as a power law in the progenitor mass ratio ($\propto \xi^{-2.01}$) for minor mergers ($\xi \la 0.1$), with a mild upturn for major mergers. These simple trends allowed us to propose a universal fitting form for the mean merger rate that is accurate to 10-20\%.

Here we go beyond the global merger rate and use the rich halo statistics in the Millennium database to quantify the merger rate as a function of halo environment, in addition to descendant mass, progenitor mass ratio, and redshift. We also investigate the environmental dependence of the progenitor (or conditional) mass function. This quantity is closely related to the merger rate and is also the most important ingredient in the EPS and excursion set models for constructing Monte Carlo merger trees.

Several recent environmental studies have used halo clustering, quantified by the halo bias, as a measure of environment (e.g. \citealt{Gottlober02, ShethTormen04,Gao2005, Harker06, JingSutoMo07, Wechsler06, GaoWhite07}). While halo bias is a powerful statistical quantity, we choose a simpler and more intuitive local environment measure and use the local mass density centred at each halo. The earlier studies that have used local overdensities as measures of halo environment have used a variety of definitions, e.g., the mean density within a sphere of some radius (ranging from 4 to $10\, h^{-1}$ Mpc) or within a spherical shell (e.g. between 2 and $5\, h^{-1}$ Mpc) \citep{LemsonKauffmann, Harker06, WangMoJing07, Maulbetsch07, Hahn08}. In this paper we compare different definitions of the local overdensity, both including and excluding the mass of the central halo itself.

In this paper we also provide an in-depth investigation of the effects of halo fragmentation on the merger rate and its environmental dependence. In FM08, we discussed how fragmentation is a generic feature of all merger trees and compared the {\it stitching} method with the conventional {\it snipping} method for handling these events. We will show here that fragmentation occurs more frequently in dense regions than in voids; understanding the effects of fragmentation on merger rates is therefore essential for obtaining robust results in dense environments. There are three general types of approaches to handling fragmentations: do nothing ({\it snipping}), {\it stitching} together fragmented haloes, or {\it splitting} up the common progenitor of the fragmented haloes. We will compare five algorithms for handling fragmentations based on these three approaches and show that except for one algorithm, all the algorithms give similar merger rates to within 20\%.

This paper is organised as follows. In \S~\ref{Definitions} we briefly review how haloes and merger trees are constructed from the particle data in the Millennium simulation. Statistics detailing the distribution of halo mass at different redshifts are summarised in Table~\ref{table:MassBins}. In \S~\ref{MeasuringEnvironment} we compare four local density measures and their distributions in relation to halo mass. Three of the measures use the dark matter mass in a sphere centred at a given halo, either including or excluding the mass of the central halo. The fourth measure is motivated by observables such as luminosity-weighted galaxy counts and uses only the masses of the haloes within a sphere. \S~\ref{MainResults} contains the main results of this paper, where we quantify how the merger rate is amplified in denser regions and suppressed in voids for redshifts $z=0$ to 2 over three decades of halo mass ($10^{12} - 5\times 10^{15} M_\odot$). A simple power-law fitting function for this environmental dependence is proposed, which can be used in combination with the fit for the global rate presented in FM08. We also show that the progenitor (or conditional) mass function has a similar environmental trend as the merger rate. Even though this is expected given that the two quantities are closely related, this result demonstrates directly that the excursion set model is incomplete. In \S~5, we present statistics of halo fragmentations, compare five algorithms for handling these events, and illustrate the robustness of the results reported in \S~4. The Appendix provides a discussion of the self-similarity of the merger rate and its environmental dependence in the context of the choice of mass and environment variables used in the fitting formula.

\section{Haloes in the Millennium Simulation} \label{Definitions} 
\begin{table*}
	\centering 
	\begin{tabular}
		{ll|ccccc} & & \multicolumn{5}{c}{Mass Percentile} \\
		Redshift & Quantity & 0-40\% & 40-70\% & 70-90\% & 90-99\% & 99-100\% \\
		\hline \hline \multirow{3}{*}{$z=\ZA\,(\dz=0.06)$} & Number of haloes & 192038 & 144028 & 96019 & 43208 & 4800\\
		& Mass bins ($10^{12} M_\odot$) & $1.2-2.1$ & $2.1-4.5$ & $4.5-14$ & $14-110$ & $>110$ \\
		& $\nu$ bins & 0.75-0.81 & 0.81-0.92 & 0.92-1.11 & 1.11-1.63 & 1.63-4.30\\
		\hline \multirow{3}{*}{$z=\ZB\,(\dz=0.06)$} & Number of haloes & 188258 & 141194 & 94129 & 42358 & 4706\\
		& Mass bins ($10^{12} M_\odot$) & $1.2-2.0$ & $2.0-4.1$ & $4.1-12$ & $12-74$ & $>74$ \\
		& $\nu$ bins & 0.95-1.03 & 1.03-1.15 & 1.15-1.37 & 1.37-1.93 & 1.93-4.66 \\
		\hline \multirow{3}{*}{$z=\ZC\,(\dz=0.09)$} & Number of haloes & 172568 & 129426 & 86284 & 38827 & 4314 \\
		& Mass bins ($10^{12} M_\odot$) & $1.2-1.9$ & $1.9-3.7$ & $3.7-9.5$ & $9.5-48$ & $>48$ \\
		& $\nu$ bins & 1.22-1.32 & 1.32-1.46 & 1.46-1.70 & 1.70-2.27 & 2.27-4.73 \\
		\hline \multirow{3}{*}{$z=\ZD\,(\dz=0.17)$} & Number of haloes & 116830 & 87622 & 58415 & 26286 & 2920 \\
		& Mass bins ($10^{12} M_\odot$) & $1.2-1.8$ & $1.8-3.0$ & $3.0-6.5$ & $6.5-24$ & $>24$ \\
		& $\nu$ bins & 1.74-1.86 & 1.86-2.01 & 2.01-2.27 & 2.27-2.85 & 2.85-5.20 \\
		\hline 
	\end{tabular}
	
	\caption{ Halo mass bins and number statistics at redshifts $z=\ZA,\ZB,\ZC$ and $\ZD$ from the Millennium simulation used in this paper. The bins are computed assuming fixed mass-percentile bins (header row). Listed are the number of haloes in each bin, and the corresponding mass and $\nu$ boundaries for each percentile bin. The highest 1\% mass bins extend out to $5.2\times10^{15},3\times10^{15},1.3\times 10^{15}$, and $4.4\times10^{14} M_\odot$ for $z=\ZA,\ZB,\ZC$ and $\ZD$ respectively. } \label{table:MassBins} 
\end{table*}

The Millennium simulation \citep{Springel05} follows the evolution of roughly $2\times 10^7$ dark matter haloes from redshift $z=127$ to $z=0$ in a $500 h^{-1}$ Mpc box using $2160^3$ particles of mass $1.2\times10^9 M_\odot$ (all masses quoted in this paper include the factor of $h^{-1}$). It assumes a $\Lambda$CDM model with $\Omega_m=0.25$, $\Omega_b=0.045$, $\Omega_\Lambda=0.75$, $h=0.73$ and an initial power-law distribution of density perturbations with index $n=1$ and normalisation $\sigma_8=0.9$.

A friends-of-friends (FOF) group finder \citep{Davis85} with a linking length of $b=0.2$ is used to identify haloes in the simulation. Each FOF halo (henceforth \emph{halo}) thus identified is further broken into constituent subhaloes (each with at least 20 particles or $2.35\times 10^{10} M_\odot$) by the SUBFIND algorithm which identifies gravitationally bound substructures within the host FOF halo (for more on SUBFIND, see \citealt{Springel01SUBFIND}).

The subhaloes are connected across the 64 available redshift outputs to form a subhalo merger tree. Mergers are complicated processes and the particles in a given subhalo will not necessarily end up in a {\it single} subhalo in the subsequent output. As such, a subhalo is chosen to be the descendent of a progenitor subhalo at an earlier output if it hosts the largest number of bound particles in the progenitor subhalo. The resulting merger tree of the subhaloes can be used to construct the merger tree of the FOF haloes, although we have discussed at length in FM08 that this construction is non-trivial due to the fragmentation of FOF haloes. Our main results reported in Sec.~4 use the \emph{stitching} tree of FM08. Since fragmentation occurs more frequently in denser environments, we provide a detailed comparison in Sec.~5 between stitching and four alternative algorithms to test the robustness of our results.

The Millennium database provides a number of mass measurements for each identified FOF halo. We use the total mass of the particles connected to an FOF by the group finder. \cite{Tinker} argue that spherical overdensity measures of mass are more closely linked to cluster observables than FOF measures and, therefore, are to be preferred. We have found, however, that the FOF mass definition is more robust in the context of merging haloes than definitions that make assumptions about halo geometry and virialization (for the simple reason that merging haloes are typically not virialized at the simulation outputs immediately preceding and following a merger event; see also \citealt{WhiteMass}).

We study the dependence of halo growth on halo environment in a variety of halo mass bins at different redshifts. Since the most massive haloes at $z=0$ are more massive than the most massive haloes at higher redshifts, we use mass bins with boundaries that vary with redshift such that each mass bin contains a fixed percentage of haloes. Table~\ref{table:MassBins} lists the five percentile bins used in our study, and the corresponding number of haloes and halo masses at $z=\ZA,\ZB,\ZC,$ and $\ZD$. We note that even in the highest 1\% mass bin, there are 4000 to 5000 cluster-mass haloes at $z\la 1$ available for this study.

Table~\ref{table:MassBins} also lists the range of $\nu$ for each mass bin, where $\nu=\delta_c(z)/\sigma(M)$ is often used as a mass variable for comparing haloes over different redshifts. Here $\sigma(M)$ is the variance of the linear density perturbations and $\delta_c(z)$ is the critical overdensity at redshift $z$, where $\delta_c(z)=1.686/D(z)$ and $D(z)$ is the linear growth function of density perturbations in $\Lambda$CDM. A comparison of the FOF mass versus $\nu$ as the mass variable is provided in the Appendix.

Our notation is as follows. When computing the merger rate, we refer to the haloes at the lower redshift as the descendants and label their masses by $M_0$. The progenitors of a given descendant halo at a (slightly) higher redshift are labelled $M_1, M_2, M_3...$, where $M_1\geq M_2\geq M_3...$ by our convention. The mass ratio of the progenitors is defined as $\xi=M_{i\geq2}/M_1$. In this paper we find that there are sufficient halo statistics from the Millennium simulation for studying the environmental dependence of the descendant haloes over a range of redshifts, and shall present results at $z_0=\ZA,\ZB,\ZC$, and $\ZD$ and their progenitors at $z_1=z_0+ \Delta z$, where $\Delta z= 0.06,0.06, 0.09$, and 0.17, respectively.

\section{Measuring Halo Environment} \label{MeasuringEnvironment} 

In this paper we quantify a halo's local environment using the local mass density centred at the halo. In this section we examine four definitions of density. Three of them are computed using the dark matter particles in a sphere of radius $R$ centred at a halo, either with or without the central region carved out (see Sec~3.1-3.3). The fourth definition is computed using the masses of only the haloes rather than all the dark matter (Sec.~3.4). This last environmental measure based on mass-weighted halo counts has the advantage that it can be linked to observables such as luminosity-weighted galaxy counts.
\begin{figure*}
	\centering 
	\includegraphics{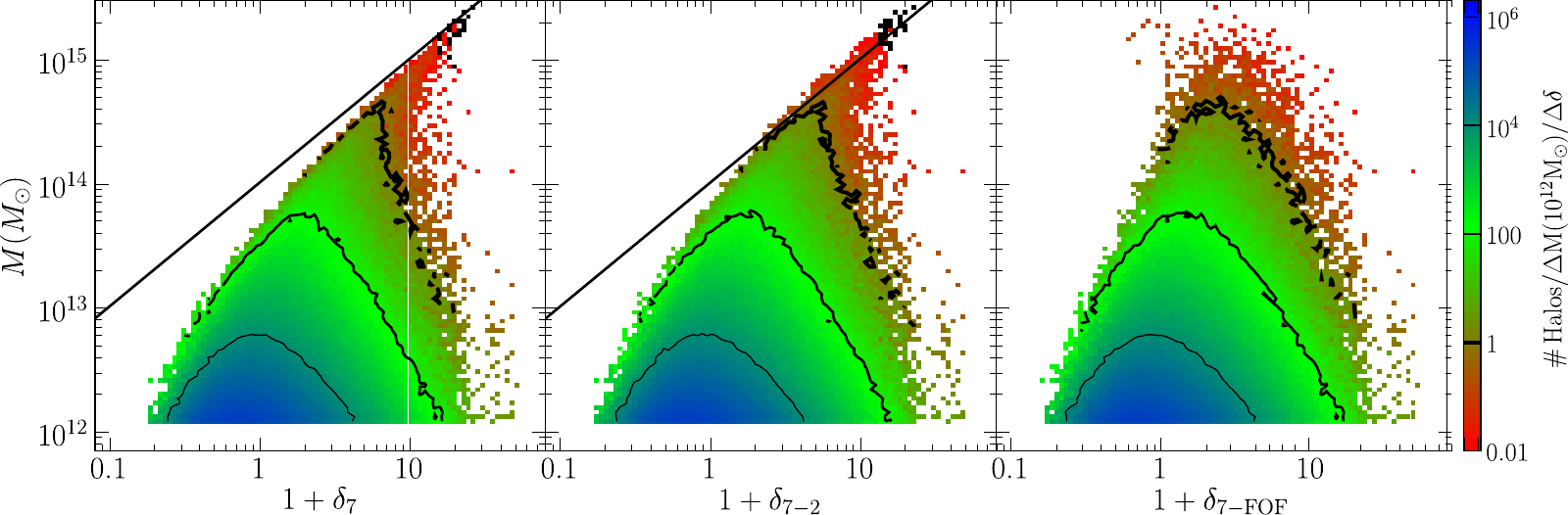} \caption{Scatter plots of halo mass vs three measures of halo environment for all FOF haloes above $1.2\times 10^{12} M_\odot$ (1000 particles or more) in the $z=0$ Millennium simulation output. The colour scale indicates the number of haloes present in each $(\delta, M)$ grid cell normalised by the bin size, and the contours are drawn at the 1, 100, and $10^4$ bin levels (decreasing line width). The left panel uses $1+\ds$, the density in a sphere of radius $7 h^{-1}$ Mpc centred at each halo. The black line is $1+\ds=M/V_7/\bar{\rho}_m$ (see text). The middle panel shows $1+\dst$, the density in a shell between $2 h^{-1}$ and $7 h^{-1}$ Mpc. The right panel uses $1+\dsfof$ by subtracting the halo mass from $\ds$. At the high mass end, the halo itself is the main contribution to $\ds$ and $\dst$, leading to the tight correlation between $\delta$ and $M$ in the upper right region in the left and middle panels. The right panels shows that this correlation is largely removed when $\dsfof$ is used, which subtracts out the FOF mass of the central halo. The variable $\dsfof$ is therefore a more independent measure of the immediate environment {\it outside} of the haloes.} \label{fig:DeltaComparison} 
\end{figure*}

\subsection{Definitions of Environment}\label{EnvDef}

Only a few studies of halo environment have used local overdensities as measures of environment \citep{LemsonKauffmann, Harker06, WangMoJing07, Hahn08}. By contrast, many studies have used the halo bias as a proxy for halo environment, which is obtained by taking the ratio of the halo-halo (or halo-mass) two-point correlation function to the underlying dark matter two-point correlation function (e.g., \citealt{Gottlober02, ShethTormen04,Gao2005, Harker06, JingSutoMo07, Wechsler06, GaoWhite07}). Typically, these studies explore the dependence of bias on a variety of tracers of the halo growth history such as formation redshift, concentration, and number of major mergers. This technique has yielded clear signs of environmental dependence, particularly when combined with the marked correlation function statistical test \citep{Gottlober02, ShethTormen04, Harker06, Wechsler06}.

The connection between a halo's local density and the halo bias, however, is not entirely straightforward. The two quantities are certainly correlated, e.g., the two-point correlation function of objects in denser regions is typically higher than that in less dense regions \citep{AbbasSheth}. However, the local density is a simple quantity that can be computed for each halo, whereas the bias is a statistical measure of clustering strength computed by averaging over a large number of pairs of haloes and particles over a range of pair separations.

The rich statistics of the Millennium simulation over large dynamic ranges in both mass and redshift make it possible to use the more intuitive local density as a measure of environment.

To compute the local overdensity in a halo's neighbourhood, we centre either a sphere or shell on the halo at spatial coordinates $\bf{x}$ and define the halo's environment by
\begin{equation}
	\delta_R({\bf x}) \equiv \frac{\rho_R({\bf x})-\bar{\rho}_m}{\bar{\rho}_m} \label{delta1} 
\end{equation}
for a sphere of radius $R$, or 
\begin{equation}
	\delta_{R_o-R_i} \equiv \frac{\delta_{R_o}R_o^3-\delta_{R_i}R_i^3}{R_o^3-R_i^3} \label{delta2} 
\end{equation}
for a shell of inner and outer radii $R_i$ and $R_o$. Here $\bar{\rho}_m$ is the mean matter density in the simulation box, and $\rho_R(\bf{x})$ is the mean density of a sphere of radius $R$ centred at $\bf{x}$. We also propose an environmental measure, $\dRfof$, computed by subtracting out the FOF mass $M$ of the central halo within a sphere of radius $R$: 
\begin{equation}
	\dRfof \equiv \delta_R - \frac{M}{V_R \bar{\rho}_m} \,, \label{delta3} 
\end{equation}
where $V_R$ is the volume of a sphere of radius $R$. Note that unlike the shell measure, this measure makes no assumption about the central halo's shape.

To compute $\rho_R(\bf{x})$, one would need all the particle positions from the Millennium simulation, which are not available on the online public database. The database, however, does provide the density on a $256^3$ cubic grid (with a grid spacing of $1.95\,h^{-1}$ Mpc) computed from the dark matter particles in the simulation using the Cloud-in-Cell (CIC) interpolation scheme. We use this data to sum up the contributions within the sphere centred at $\bf{x}$ to evaluate $\rho_R(\bf{x})$. We note that the grid in the database is indexed using a Peano-Hilbert space filling curve, which we have mapped to spatial coordinates in order to compute $\rho_R$.

It is important to choose an appropriate radius $R$ (or $R_o-R_i$) in equations~(\ref{delta1})-(\ref{delta3}) when computing halo environment. \cite{LemsonKauffmann} used both $\delta_{10}$ and $\delta_{5-2}$ (in units of $h^{-1}$ Mpc) but failed to detect any environmental dependence in the formation redshift for haloes with masses between $2\times10^{12}$ and $10^{14} h^{-1} M_\odot$. \cite{Harker06}, following \cite{LemsonKauffmann}, used $\delta_{5-2}$ and \emph{did} detect environmental dependence in Millennium for haloes with masses between $2\times10^{12}$ and $10^{14} h^{-1}M_\odot$. Similarly, \cite{Hahn08} used $\delta_{5}$ and $\delta_{5-2}$ and found environmental dependence for haloes with masses between $2\times 10^{10}$ and $1.6\times 10^{11} h^{-1} M_\odot$. We will show in \S\ref{DeltaDistribution} that $R=7 h^{-1}$ Mpc is an adequate choice that effectively characterises the environments of massive halos.

\subsection{Disentangling Environment and Mass} \label{Detangling}

Since the goal of this paper is to quantify the dependence of merger rates on halo environment as well as mass, it is essential for us to first examine the extent to which these two variables are independent measures of halo properties. This is particularly relevant considering that our measure of environment is based on the local mass density. We note that in the literature mass is often used loosely to refer to environment, e.g., clusters are considered denser environments than galaxies. This interpretation is valid for galaxy counts. We are concerned with FOF haloes (and not subhaloes or galaxies) here, however. As we will show, haloes of all masses can reside in a wide range of overdensities.

To study the relation between halo mass and environment, we present a scatter plot of the mass of every FOF halo (above 1000 particles $M>\MMIN M_\odot$) at $z=0$ in the Millennium simulation versus its local $1+\delta$ in Fig.~\ref{fig:DeltaComparison}. Three definitions of local density are shown for comparison: all mass within a $7 h^{-1}$ Mpc sphere ($\ds$; left panel), all mass within $7 h^{-1}$ Mpc excluding the central $2 h^{-1}$ Mpc ($\dst$; middle panel), and all mass within $7 h^{-1}$ Mpc excluding the central FOF mass ($\dsfof$; right panel).
\begin{figure*}
	\centering 
	\includegraphics{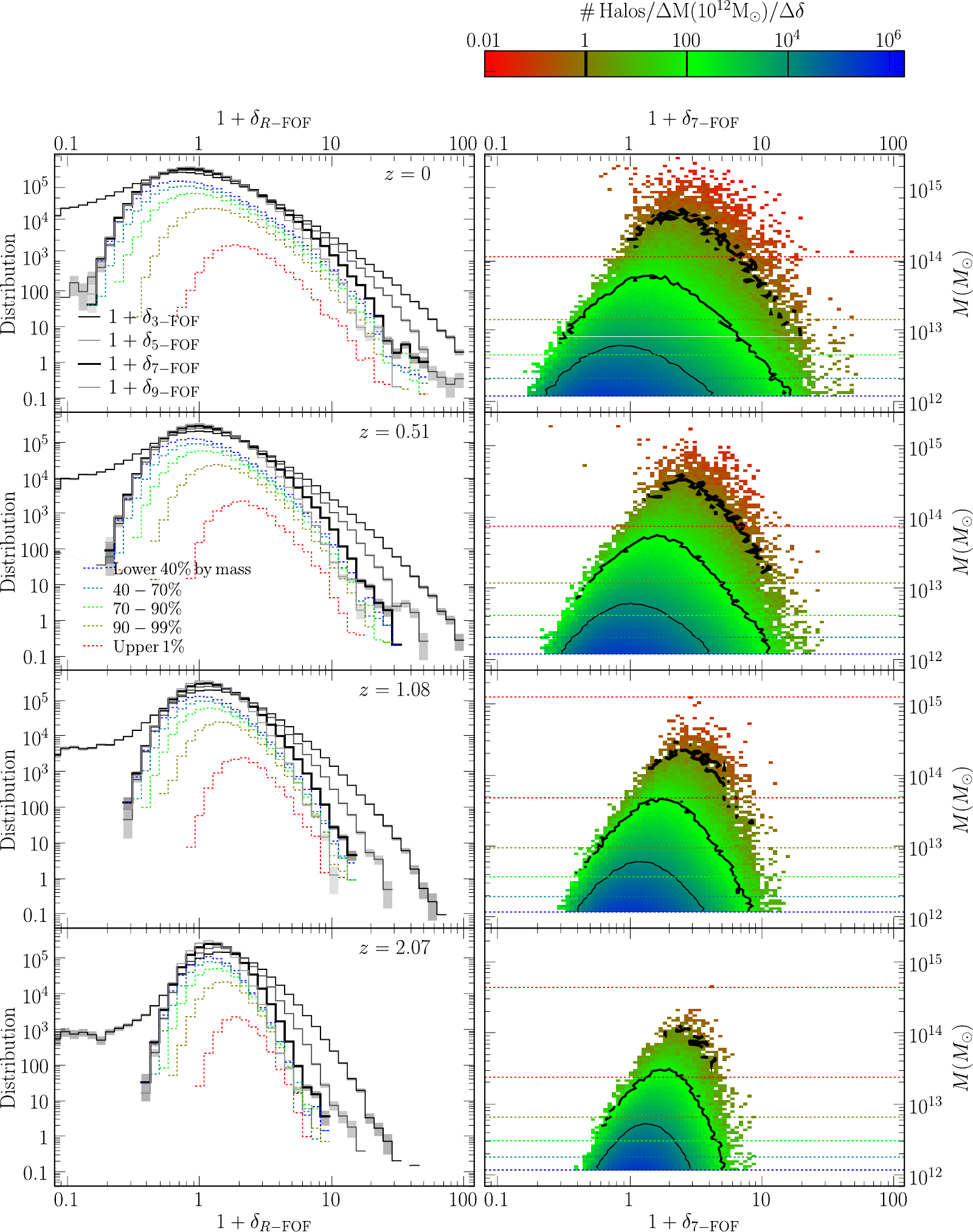} \caption{Left panels: Distribution of the environmental variable $1+\delta_{R-FOF}$ defined in equation~(\ref{delta3}) at four redshifts $z=\ZA, \ZB, \ZC$ and $\ZD$ (top to bottom) computed from all haloes with $M>\MMIN$ (i.e. above 1000 particles) in the Millennium simulation. The broadening of the distribution with decreasing $z$ illustrates the effect of gravitational instability. Within each panel, the four grey-scale histograms compare four smoothing radii $R$ in $h^{-1}$ Mpc: 3 (thin black), 5 (thin dark grey), 7 (thick black) and 9 (thin light grey); the five coloured dotted histograms compare the separate contributions to the $\delta_{7-FOF}$ distribution from haloes of different mass percentile bins: top 1\% (red), 90-99\% (olive), 70-90\% (green) 40-70\% (cyan), and bottom 40\% (blue). Right panels: Similar scatter plot as the right panel of Fig.~\ref{fig:DeltaComparison} but at four redshifts. The horizontal dotted lines mark the five mass percentile bins used in the left panels and listed in Table~\ref{table:MassBins}.} \label{fig:DeltaDistribution} 
\end{figure*}

Fig.~\ref{fig:DeltaComparison} shows that galaxy-size haloes ($\sim 10^{12} M_\odot$) reside in a wide range of environmental densities from extreme underdense regions of $\delta \sim -0.8$ to regions with $\delta > 20$. The three panels show similar distributions of $\delta$ for these low mass haloes regardless of the definition of $\delta$ used. The high mass haloes, on the other hand, have very different distributions of $\delta$. The rich statistics of the Millennium simulation allow us to study halo mass out to $2\times 10^{15} M_\odot$, an order of magnitude higher than in previous studies. At $5\times 10^{14} M_\odot$ and above, the spherical and shell measures of overdensity, $\ds$ and $\dst$, are seen to be tightly correlated with the halo mass (left and middle panels). In addition, all the points lie close to the line that represents the density in a $7 h^{-1}$ Mpc sphere computed from the FOF halo mass {\it alone}, that is, $M/V_7/\bar{\rho}_m$, where $V_7$ is the volume of a sphere of radius $7 h^{-1}$ Mpc. This trend clearly indicates that the central haloes are dominating the local overdensity at masses above $\sim 5\times 10^{14} M_\odot$, and both $\ds$ and $\dst$ are tracing the central halo mass rather than the overdensities in the neighbourhood {\it outside} the virial radius of the halo. Even though $\dst$ subtracts out the central $2 h^{-1}$ Mpc regions, the tight residual correlation seen in the middle panel suggests that this quantity does not cleanly remove the contribution made by the central halo, probably because these massive haloes extend well beyond $2 h^{-1}$ Mpc. We have also tested $\delta_{5-2}$, the measure of environment used in \citet{LemsonKauffmann, Harker06, Hahn}, and found nearly identical results as $\dst$. This correlation may not be problematic for the results reported in these earlier papers, however, as these studies did not report results beyond $\sim 10^{14} M_\odot$.

The right panel of Fig.~\ref{fig:DeltaComparison} shows that our third environmental variable, $\dsfof$, in equation~(\ref{delta3}) is capable of disentangling the tight correlation between halo mass and density seen for $\ds$ and $\dst$; $\dsfof$ is therefore a more robust measure of the environment {\it outside} of a halo's virial radius. It should, however, be kept in mind that haloes of different masses residing in the same 7 Mpc region will have the same $\ds$ but different $\dsfof$. A cluster-sized halo, for instance, will have a smaller value of $\dsfof$ than a neighbouring galaxy-sized halo, and the difference between the two values of $\dsfof$ will be the difference between the mass of the cluster and the galaxy (appropriately normalised). This caveat should be considered when interpreting values of $\dsfof$ across different mass bins. The spherical measure $\ds$, on the other hand, is simpler in this context. For this reason, we will report results using both $\dsfof$ and $\ds$ below.

\subsection{$\delta$ Distributions} \label{DeltaDistribution}

To gain further insight into the properties of the environmental measure $\dRfof$ of equation~(\ref{delta3}), we plot in the left panels of Fig.~\ref{fig:DeltaDistribution} the distribution of $1+\dRfof$ centred at each halo for all haloes in the Millennium database with more than 1000 particles ($M>\MMIN M_\odot$) at four redshifts $z=\ZA,\ZB,\ZC,$ and $\ZD$ (top to bottom). Within each panel, four choices of radii, $R = 3,5,7$, and $9\,h^{-1}$ Mpc, are shown for comparison (thin black, thin dark grey, thick black, and thin light grey). A comparison of the four left panels shows that the width of the $1+\dRfof$ distribution becomes broader towards lower redshifts. This is a natural consequence of gravitational instability: denser regions become denser and vice versa as the universe evolves. We will explore the implications of this effect further in the appendix.

At a given redshift, as expected for a $\Lambda$CDM model, the overdensities computed using a larger smoothing radius $R$ are generally smaller than those computed using a smaller $R$. In the voids, the distribution of $1+\delta_{3-{\rm FOF}}$ is seen to have a low $\delta$ tail that extends down to unphysical (negative) $1+\delta$; a faint remnant of this tail is also visible in $1+\delta_{5-{\rm FOF}}$ at $z=0$. This tail is due to a number of cluster-size haloes whose FOF member particles extend beyond 3 to 5 Mpc. We note that the virial radii of even the most massive halos ($10^{15}M_\odot$) do not extend beyond 3 Mpc; however we have found that the distance between the centre of an FOF's most massive subhalo and the furthest subhalo associated with said FOF can extend beyond 5 Mpc even for halos with masses of a few $\times10^{14} M_\odot$. We therefore use $1+\dsfof$ throughout this paper in order to better sample the environment surrounding these large haloes.

We break down the haloes represented by the thick black $1+\dsfof$ curve in the left panels of Fig.~\ref{fig:DeltaDistribution} into different mass bins and plot their $\delta$-distributions using dotted colour histograms. Less massive haloes cover a broader range of $1+\dsfof$ than more massive haloes, and their distribution peaks at a lower value of $\dsfof$. We note that even though the histograms for both the lower mass haloes and the total distribution at $z=0$ are peaked at a slightly negative value of $\delta$, the mean value is in fact positive, e.g., $<\ds>=0.864$ and $<\dsfof>=0.849$ for the lowest mass bin, and $<\ds>=1.04$ and $<\dsfof>=0.956$ for all the halos. The value of $<\dsfof>$ is only slightly smaller than $<\ds>$ because subtracting the FOF mass of the low mass haloes (which dominate the total distribution) makes little difference when $\delta$ is averaged over a sphere of radius as large as $7 h^{-1}$ Mpc. The mean of $\delta$ is not zero here because the overdensities are not randomly sampled but are instead centred on haloes.

The right panels in Fig.~\ref{fig:DeltaDistribution} are scatter plots of each halo's local density $1+\dsfof$ versus its mass at $z=\ZA, \ZB, \ZC$, and $\ZD$ (top to bottom). (The top panel is a repeat of the right panel of Fig.~\ref{fig:DeltaComparison}.) The mass bins based on percentiles from Table~\ref{table:MassBins} are marked by the horizontal lines. At high $z$ the haloes cover a narrower range in both $\delta$ and $M$, but the tight correlation seen in Fig.~\ref{fig:DeltaComparison} between the mass of the massive haloes and their local densities $\ds$ and $\dst$ is removed at all redshifts when $\dsfof$ is used.

\subsection{Computing $\delta$ via Halo Counts} \label{DeltaHalo}
\begin{figure}
	\centering 
	\includegraphics{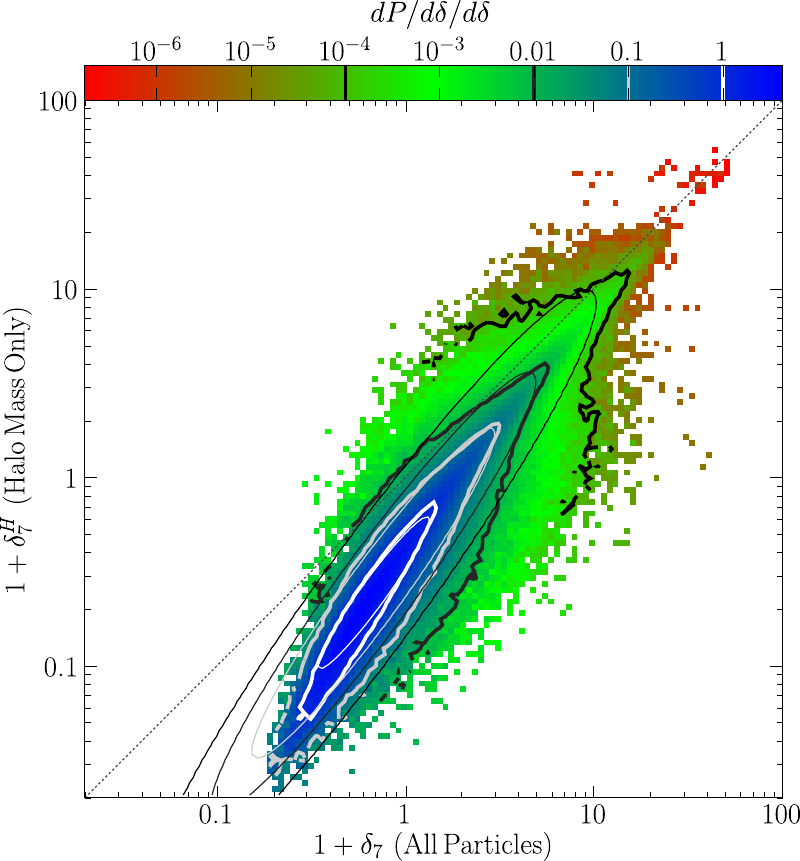} \caption{Scatter plot of two density variables: $1+\ds$ (see Fig.~1) computed from all dark matter particles centred within a $7 h^{-1}$ Mpc sphere of each halo, and the mass-weighted halo counts $1+\delta^H_7$ computed by including only masses in haloes in the same sphere (down to halo mass of $\MMIN M_\odot$). The contours are plotted at probability values of $10^{-4}$ (black), 0.01, 0.1 and 1 (white); the thick contours are for simulation data, the thin contours are from the fit in eq.~(\ref{eq:LogNormProb}). The grey dotted line is for $\ds=\delta^H_7$.} \label{fig:D7vsDH} 
\end{figure}

The local environmental measure $\ds$ is convenient from a theoretical standpoint but is not easy to measure observationally as it demands accurate knowledge of the background dark matter distribution within a large (7 Mpc) radius of the halo in question. Here we consider a more observer-friendly quantity based on the mass-weighted halo counts (above a certain mass threshold): 
\begin{equation}
	1+ \delta^H_7({\bf x}) \equiv \frac{\sum M_{\rm halo}}{V_7 \bar{\rho}_m} \label{deltaH} 
\end{equation}
where the sum is over all halos within a $7 h^{-1}$ Mpc sphere centred at $\bf x$ above some minimum mass (we use 40 particles, or $4.7\times10^{10} M_\odot$), and $V_7$ is the volume of a sphere of radius $7 h^{-1}$ Mpc. For a given halo in the Millennium simulation, we compute this quantity by summing over all haloes whose \emph{centres} lie within the $7 h^{-1}$ Mpc sphere centred on the halo in question. We do not account for the fact that halos near the boundary may only strictly contribute a fraction of their mass to the $7 h^{-1}$ Mpc sphere.

The resulting mass-weighed halo counts $1 + \delta_7^H$ is plotted against $1+\delta_7$ computed from the CIC density grid in Fig.~\ref{fig:D7vsDH}. The 2d-histogram is normalised to have unit area and can be thought of as a bivariate probably distribution. We note that, while $\delta_7$ is generally greater than $\delta_7^H$ as expected, there are regions with $\delta_7^H>\delta_7$, particularly in dense environments. This is due to the fact that a halo's entire mass contributes to $\delta_7^H$ if its centre lies within the $7 h^{-1}$ Mpc sphere in question.

We approximate the distribution with a two-dimensional log-normal distribution. Since the variables are correlated, the fitting form has five parameters and is given by 
\begin{equation}
	\frac{dP}{d\delta_7 d\delta^H_7}=\frac{1}{2\pi x_1 x_2 \sigma_1 \sigma_2} \exp\left[-\frac{\ln(x_1)^2}{2\sigma_1^2}--\frac{\ln(x_2)^2}{2\sigma_2^2}\right] \,, \label{eq:LogNormProb} 
\end{equation}
where $\ln(x_1)$ and $\ln(x_2)$ are uncorrelated variables that are simply linear combinations of $\ln(1+\delta_7)$ and $\ln(1+\delta_7^H)$ given by 
\begin{equation}
	\left[ 
	\begin{array}{c}
		\ln(x_1)\\
		\ln(x_2) 
	\end{array}
	\right]=\left[ 
	\begin{array}{cc}
		\cos\theta & \sin\theta\\
		-\sin\theta & \cos\theta 
	\end{array}
	\right]\left[ 
	\begin{array}{c}
		\ln(1+\delta_7)-\mu_{1}\\
		\ln(1+\delta_7^H)-\mu_{2} 
	\end{array}
	\right] \,. 
\end{equation}
Here $\mu_{1}$ and $\mu_{2}$ denote the mean values of $\ln(1+\delta_7)$ and $\ln(1+\delta_7^H)$ respectively, $\theta$ is an angle that quantifies the correlation between the two $\delta$s, and $\sigma_1$ and $\sigma_2$ are the standard deviations along the major and minor axes defined by $\theta$. The best fit values for these five parameters are $\mu_1=0.210$, $\mu_2=-0.549$, $\sigma_1=1.03$, $\sigma_2=0.141$, $\theta=0.943$. The thin contours in Fig.~\ref{fig:D7vsDH} represent the resulting fit.

We have also computed a simpler power-law fit that can be used to approximate the mean relation between the two densities: 
\begin{equation}
	\ln(1+\delta_7^H)=1.28\ln(1+\delta_7)-0.865 \,. 
\end{equation}
Both fitting forms can be used to convert back and forth between $\delta_7$ and $\delta_7^H$.

\section{Environmental Dependence} \label{MainResults}

\subsection{Halo Merger Rate} \label{HMR}
\begin{figure*}
	\centering 
	\includegraphics{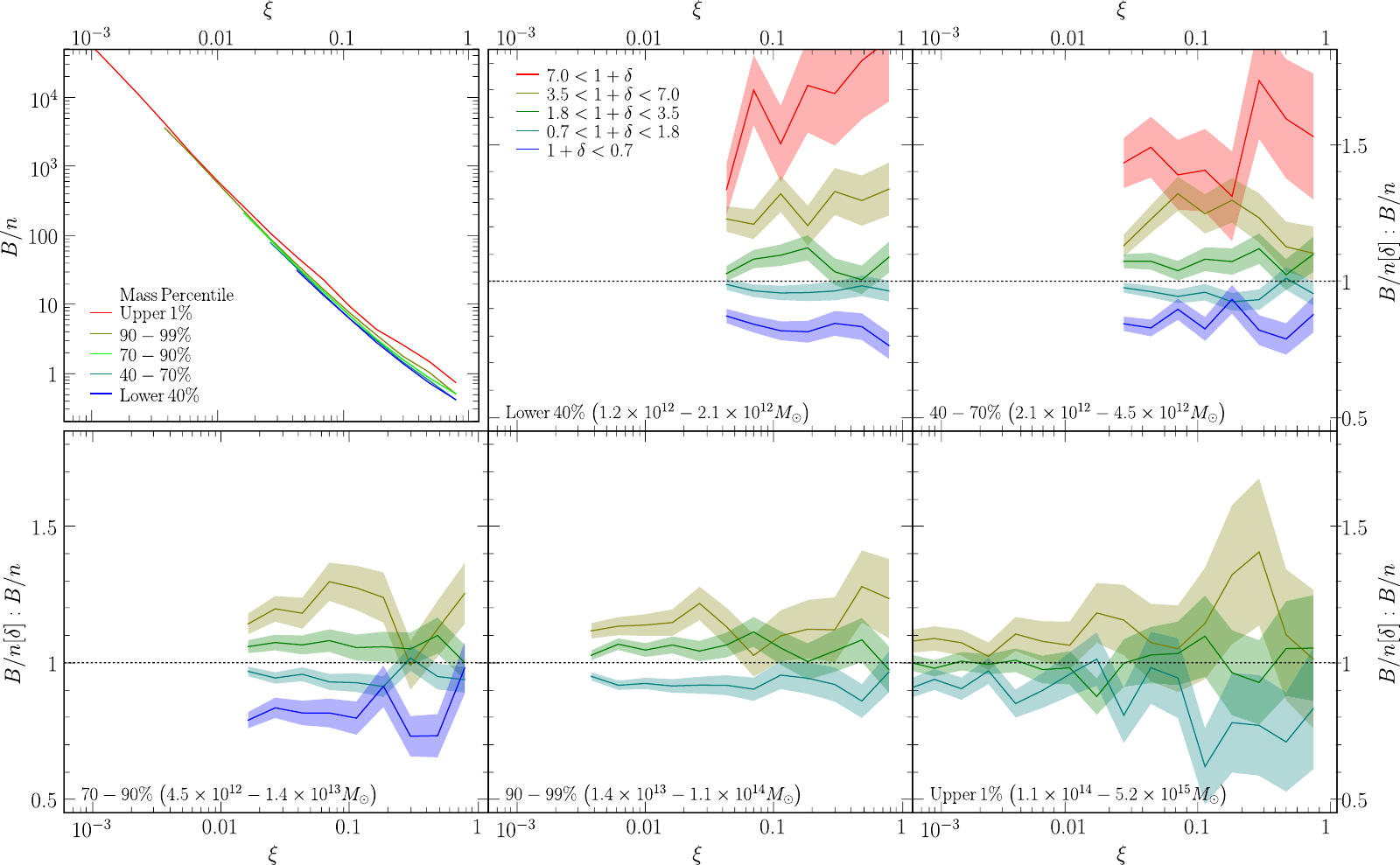} \caption{Halo merger rate and its environmental dependence on the local overdensity $\dsfof$ measured in a $7 h^{-1}$ Mpc sphere excluding the central FOF halo mass. Top left panel: The global mean merger rate $B/n$ (in units of mergers per descendant halo per unit redshift per $\xi$ bin) as a function of the progenitor mass ratio $\xi$ for descendant haloes in five mass percentile bins (see Table~\ref{table:MassBins}). The results are computed using the $z=0$ and 0.06 outputs from the Millennium simulation. The higher mass curves extend down to lower $\xi$ because we have chosen a fixed minimal progenitor mass (40 particles) for all descendants. Other five panels: The ratio of the merger rate of haloes in a given environmental bin $B/n[\delta]$ to the global mean $B/n$ as a function of $\xi$. Each panel is for a mass bin shown in the upper left panel. Within each panel, different colours show different $1+\dsfof$ bins (red for the densest regions; blue for the void regions), and the bands indicate the size of the Poisson errors. Note that the lower panels for the higher mass haloes have fewer $\delta$ curves since $\dsfof$ for these haloes spans a narrower range. This figure clearly shows that the merger rate is higher in dense regions and lower in voids for all halo masses, and the boost or reduction factor is nearly independent of the progenitor mass ratios $\xi$. } \label{fig:Bn} 
\end{figure*}
\begin{figure*}
	\centering 
	\includegraphics{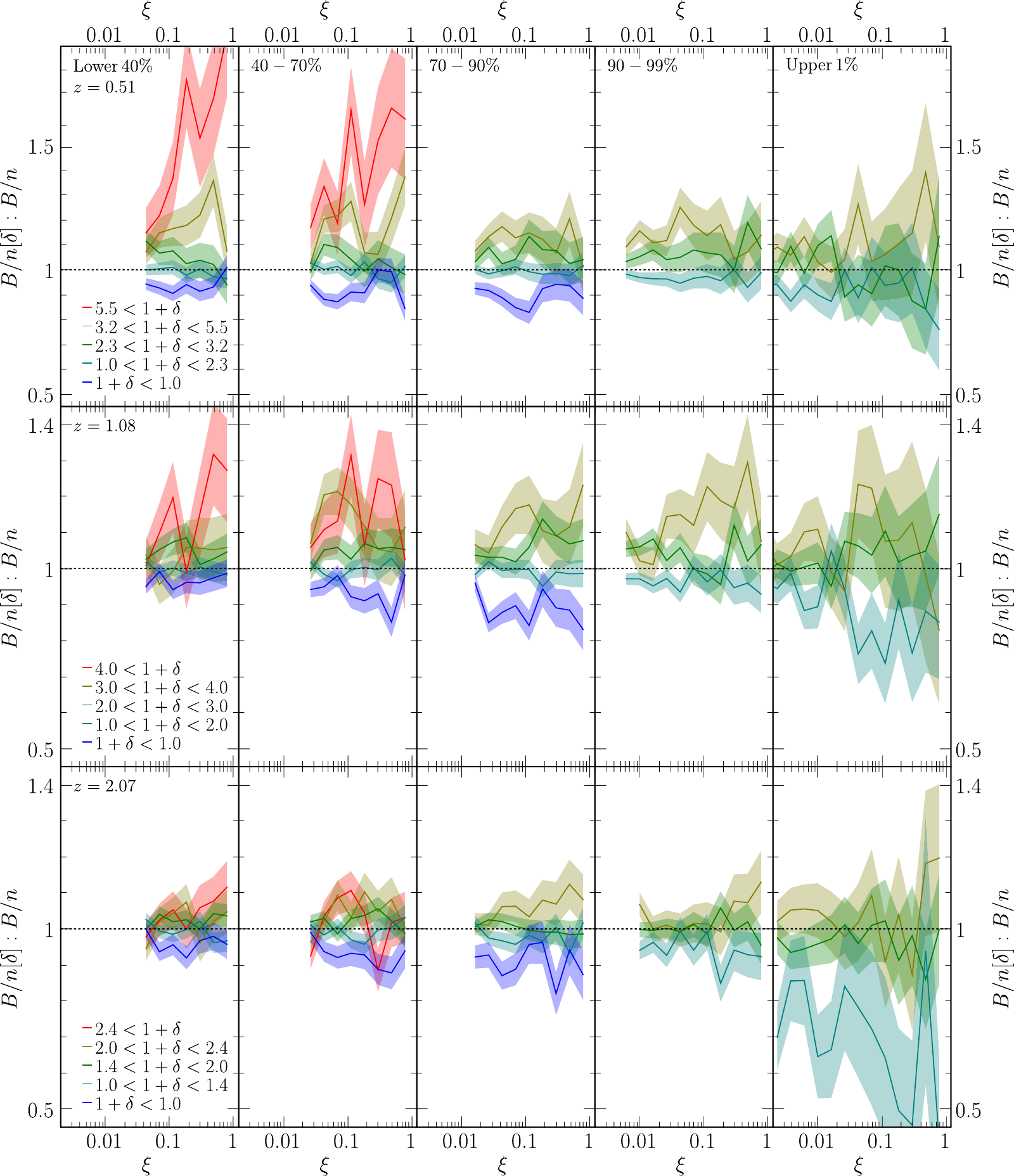} \caption{Same as the merger rate ratio plots in Fig.~\ref{fig:Bn} except at higher redshifts: $z=\ZB,\ZC,\ZD$ (from top to bottom). The five columns correspond to the five mass percentile bins (see Table~\ref{table:MassBins}). Within each panel, the coloured curves show different $1+\dsfof$ bins (red for the densest regions; blue for the void regions), and the bands indicate the size of the Poisson errors. Note that since the distribution of $\dsfof$ evolves with $z$, the corresponding $\delta$ bins (labelled in the leftmost columns) for the five coloured curves change with redshift. This figure shows that the positive correlation of the merger rate with $\dsfof$ persists out to $z\approx 2$.} \label{fig:BnHighZ} 
\end{figure*}
\begin{figure*}
	\centering 
	\includegraphics{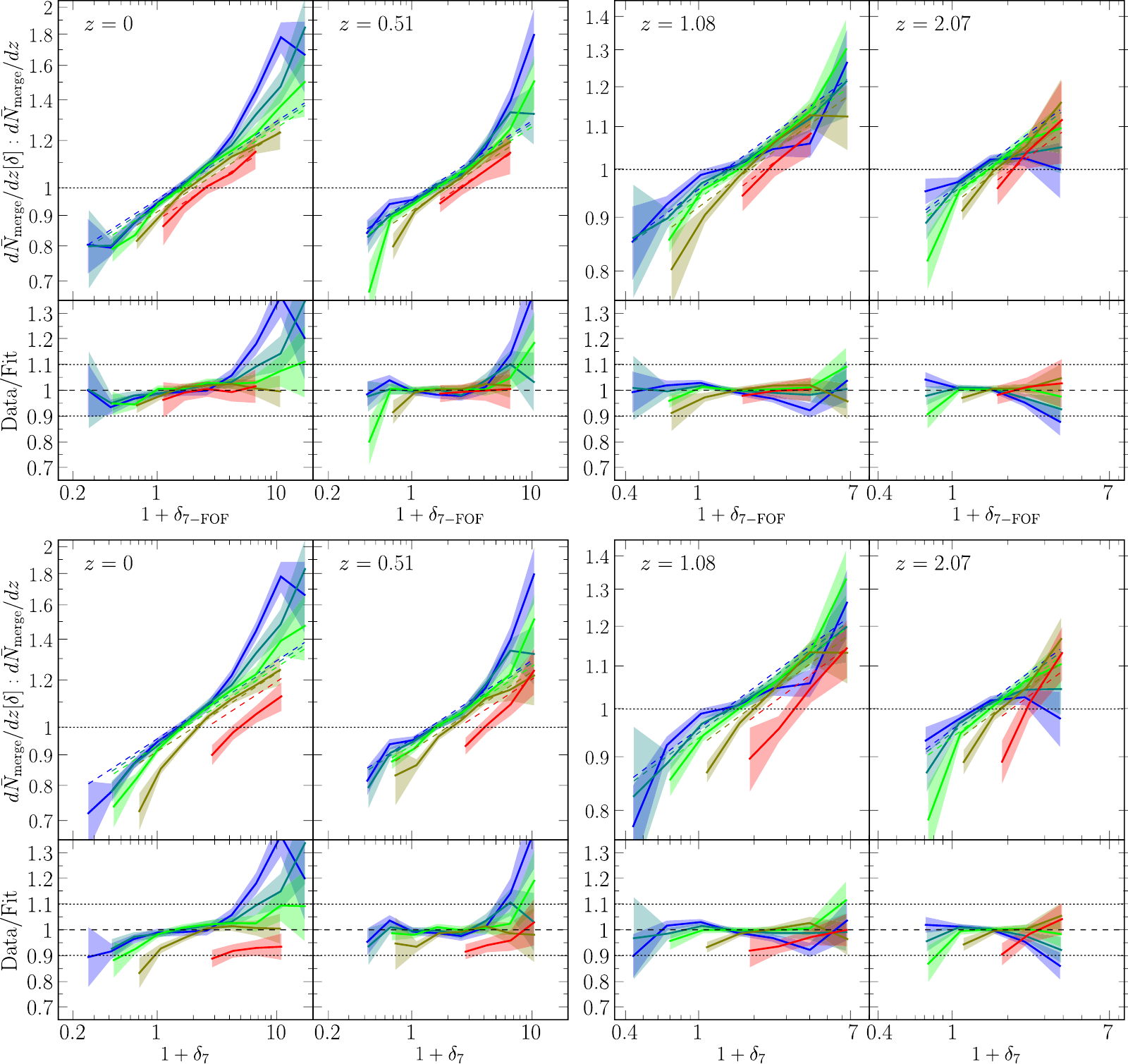} \caption{Dependence of the mean merger rate $\Rz$ ($=\int B/n d\xi$) on environmental variables $1+\dsfof$ (top figure) and $1+\ds$ (bottom figure) at four redshifts $z = \ZA,\ZB,\ZC$,and $\ZD$ (left to right). Within each figure, the top panel shows the ratio of the mean merger rate $\Rz[\delta]$ for haloes in a given environment to the global mean merger rate $\Rz$. The bottom panel plots the ratio of the simulation results to the fits, showing that eq.~(\ref{eqn:FIT}) is generally accurate to within 10\% (indicated by the dotted horizontal line). The colours correspond to the five mass percentile bins in Table~\ref{table:MassBins} (red for the highest and blue for the lowest mass bin); the bands correspond to Poisson errors. This figure shows that the positive correlation of the merger rate with environmental density is present at all mass and redshift ranges probed. } \label{fig:Bn2} 
\end{figure*}
\begin{figure*}
	\centering 
	\includegraphics{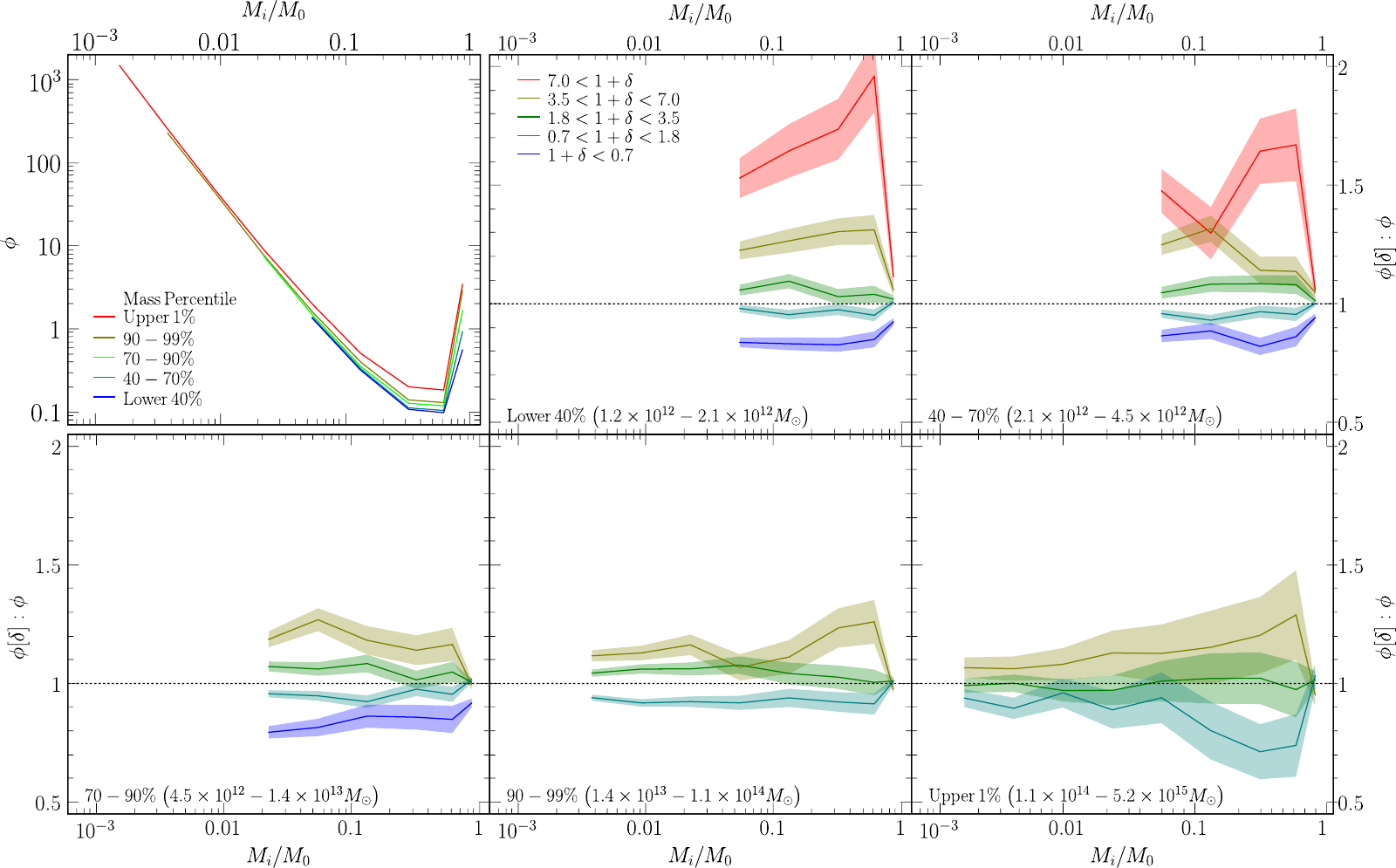} \caption{Same as Fig.~\ref{fig:Bn} except the quantity shown is the progenitor (or conditional) mass function $\phi(M_1,z_1|M_0,z_0)$ instead of the merger rate $B/n$. The results are computed using the $(z_0,z_1)=(0,0.06)$ outputs from the Millennium simulation. Similar to Fig.~\ref{fig:Bn}, we find that descendant haloes in the densest regions (red curves) have significantly more progenitors (by a factor of 1.5 to 2) than the global distribution of progenitors, whereas those in the voids (blue curves) have $\sim 20$\% to 30\% fewer progenitors.} \label{fig:CMF} 
\end{figure*}

In FM08 we defined and computed the merger rate $B/n$ as a function of progenitor mass ratio $\xi\equiv M_i/M_1$ (with $i\geq2$), descendant mass $M_0$, and redshift $z$. The rate $B/n$ is dimensionless and measures the mean number of mergers per halo per redshift interval per mass ratio. We found that in these units, the merger rate has a remarkably simple form and depends only weakly on mass and redshift. We proposed the fitting form
\begin{equation}
	\frac{B(M_0,\xi,z)}{n(M_0,z)} = A\left (\frac{M_0}{\tilde{M}}\right)^{\alpha} \xi^{\beta} \exp\left[\left(\frac{\xi}{\tilde{\xi}}\right)^{\gamma}\right] \left(\frac{d \delta_c}{dz}\right)^\eta \,, \label{eqn:Bnfit} 
\end{equation}
where $(\alpha,\beta,\gamma,\eta)=(0.083,-2.01,0.409,0.371)$, $A=0.0289$, $\tilde{\xi}=0.098$, $\tilde{M}=1.2\times10^{12}M_{\odot}$, and $\delta_c(z)\propto 1/D(z)$ is the standard density threshold normalised to $\delta_c=1.686$ at $z=0$, with $D(z)$ being the linear growth factor. This fit is accurate to 10-20\% over the mass range $10^{12}-10^{15} M_\odot$ and redshift range $z<6$.

The merger rates $B/n$ at $z=0$ for the five descendant halo mass bins in Table~\ref{table:MassBins} are reproduced for reference in the top left panel of Fig.~\ref{fig:Bn}. As shown in FM08 and indicated by equation~(\ref{eqn:Bnfit}), the merger rate is approximately a power law in $\xi$ in the minor merger regime and has a slight upturn in the major merger regime ($\xi \ga 0.2$). All five curves are nearly on top of one another, reflecting the very mild mass dependence ($\alpha \sim 0.1$) in equation~(\ref{eqn:Bnfit}). We compute each curve by first selecting the descendant haloes in a given mass bin and computing the mass ratios $\xi$ for the progenitors of these haloes. We then compute $B/n$ by counting $B$, the number of progenitors ($M_i$ with $i>2$) that lie in a given mass ratio bin ($\xi$), and dividing by $n$, the total number of descendants in the mass bin in question. See FM08 for further details of this procedure and discussions of the results.

Equation~(\ref{eqn:Bnfit}) gives the global mean merger rate averaged over all halo environments. To investigate the correlation of $B/n$ with environment, we divide each mass bin shown in the top left panel of Fig.~\ref{fig:Bn} into five environmental bins and compute $B/n[\delta]$ using $B$ {\it and} $n$ in the given $\dsfof$ bin. The remaining five panels in Fig.~\ref{fig:Bn} show our results for the ratios of $B/n[\delta]$ to the global mean $B/n$, as a function of $\xi$, for each of the five descendant mass bins. Within each panel, the different curves are for different $\dsfof$ bins for which there are sufficient halo statistics.

For descendant haloes of mass $10^{12}$ to $10^{13}M_\odot$, Fig.~\ref{fig:Bn} shows a strong environmental effect with a {\it positive} correlation between merger rates and local density: haloes in the densest regions ($1+\dsfof > 7$; red curves) experience 1.5 to 2 times more mergers than the average, while haloes in underdense regions ($1+\dsfof < 0.7$; blue curves) experience fewer mergers (by a factor of 0.7 to 0.8) than average. For group and cluster scale haloes ($10^{13}$ to $5\times 10^{15}M_\odot$) in the bottom panels, only three curves are shown for the middle three $\dsfof$ bins because massive haloes span a smaller range of $\dsfof$, as shown in Figs.~1 and 2. For each $\dsfof$ bin, the value of the merger rate ratio is quite similar for all five panels, indicating that the environmental effect, as measured by $\dsfof$, depends very weakly on halo mass. We will quantify this statement using a fitting formula below.

Fig.~\ref{fig:BnHighZ} presents the same information as Fig.~\ref{fig:Bn} but at higher redshifts ($z=\ZB, \ZC,$ and $\ZD$ from top to bottom panels). The $\delta$-bins for which the curves are too noisy are excluded. Since the distribution of $1+\dsfof$ narrows with increasing $z$, the $\dsfof$ bins span a smaller range at $z=2$ than at $z=0$. Nonetheless, we see that the environmental dependence observed at $z=0$ persists out to $z=2$. Moreover, haloes with similar $1+\dsfof$ experience similar amplifications or reductions in the merger rate regardless of mass and redshift.

An additional feature to note in Figs.~\ref{fig:Bn} and \ref{fig:BnHighZ} is that the curves are horizontal: environmental effect is therefore largely independent of the mass ratio $\xi$; that is, major and minor merger rates are boosted or dampened by a halo's environment by a similar factor. We can therefore integrate over the mass ratio parameter without diluting the environmental effect: 
\begin{equation}
	\Rzf(M,z) = \int_{\ximin}^1 \frac{B(M,\xi,z)}{n(M,z)} d\xi 
\end{equation}
where $\Rz$ is the mean merger rate per unit redshift per descendant halo with progenitor mass ratio above $\ximin$.

The value of $\Rz$ clearly depends on $\ximin$ and is larger when more minor mergers are included (see, e.g., Figs.~7 and 8 of FM08). For a fixed resolution mass (our choice is 40 particles or more for progenitor haloes), $\ximin$ extends down to lower values for higher mass descendants. For a fair comparison across halo mass bins, one should in principle use a fixed $\ximin$ for all mass bins at the expense of throwing out resolved progenitors for high mass descendant haloes. Since we plot ratios of the merger rates, however, the fact that more massive haloes are better resolved and have higher $\Rz$ is normalised out. It is therefore possible to make a fair comparison across mass bins without throwing out any resolved progenitors.

Fig.~\ref{fig:Bn2} shows the ratio of the merger rate $\Rz$ as a function of $1+\delta$ at four redshifts ($z = \ZA,\ZB,\ZC,\ZD$ from left to right). For comparison, the results for two environmental measures are included: $\dsfof$ (top figure) and $\ds$ (bottom figure). Within each figure, the upper panel shows the simulation data and the lower panel compares the data to the analytic fitting formula discussed below. The five curves in each panel are for the five mass bins listed in Table~\ref{table:MassBins}. This figure shows the same trend as Fig.~\ref{fig:Bn}: haloes in the densest regions at $z=0$ experience up to $\sim 1.5$ times as many mergers as the average halo, whereas the merger rate in the voids is 20 to 30\% below the global average.

To quantify the dependence of the merger rate on $\delta$, we introduce 
\begin{equation}
	\Rzf (\delta,M,z) \approx \Rzf (M,z)\times f(\delta,M,z) \,, \label{eqn:dNdz} 
\end{equation}
where we have made use of the fact that the environmental dependence is independent of $\xi$ to define $f$, and $dN/dz(M,z)$ is the global merger rate from FM08. We provide two fitting forms for $f$ using $\delta=\dsfof$ and $\ds$, respectively. We find that a simple power-law and redshift-independent form works well: 
\begin{eqnarray}
	&& f(\dsfof,M) = 0.963\, (1+\dsfof)^{0.130} \left(\frac{M}{10^{12} M_\odot}\right)^{-0.0156} \nonumber\\
	&& f(\ds,M) = 0.968\, (1+\ds)^{0.135} \left(\frac{M}{10^{12} M_\odot}\right)^{-0.0252} \,. \label{eqn:FIT} 
\end{eqnarray}
The fits are performed using data from all four redshifts simultaneously (over much finer mass bins than those shown in Fig.~\ref{fig:Bn2}). The reduced $\chi^2_\nu$ for the two fits is 0.95 and 1.08, respectively. Errors are computed assuming Poisson statistics and are represented by the filled regions in Fig.~\ref{fig:Bn2}. The resulting fit is shown as dashed curves in the upper row of each figure, and the ratio of the simulation data to the fits is shown in the lower rows. The fits are seen to be accurate to within 10\% for a wide range of $\delta$ and $M$, except for low mass haloes with $\delta\ga 5$ at $z=\ZA$ and $\ZB$, where the rates steepen suddenly.

As we will discuss in \S~\ref{Split}, our extensive tests using various algorithms suggest that the merger rate in this particular parameter range (i.e. low mass, low $z$, high density) depends sensitively on the post-processing algorithm used to handle fragmentations in the merger tree, and variations of order 20\% or more among different algorithms are observed. We therefore do not attempt to use a fitting form more complicated than equation~(\ref{eqn:FIT}) to get a better fit in this uncertain regime.

It is interesting to note that when $\ds$ is used, instead of $\dsfof$, as the environment variable, the only change in the fit in equation~(\ref{eqn:FIT}) is a stronger dependence on halo mass. This trend makes sense since the difference between $\ds$ and $\dsfof$ is $\ds - \dsfof = M/(V_7 \bar{\rho}_m)$ (see eq.~[\ref{delta3}]). This difference is negligible for galaxy-scale haloes (e.g. $\ds-\dsfof\sim 0.01$ for $10^{12} M_\odot$) but becomes larger for more massive haloes, reaching $\ds-\dsfof \sim 10$ at $M\sim 10^{15} M_\odot$. The five curves for the five mass bins at a given $z$ in Fig.~\ref{fig:Bn2} are therefore more spread out when $\ds$ is used as the variable, resulting in a stronger mass dependence.

We have chosen to use halo mass and local density as variables in equation~(\ref{eqn:FIT}). It is interesting to ask if other choices of variables may lead to a more accurate fit across the wide ranges of halo masses, densities, and redshifts shown in Fig.~\ref{fig:Bn2}. For instance, the variance of the linear density perturbation $\sigma(M)$ and the scaled density threshold $\nu(M,z) = \delta_c/\sigma(M) D(z)$ are commonly used to characterise mass and redshift dependence of halo properties (e.g., the mass function). We test these variables and describe the results in appendix~\ref{AppendixSelfSim}. Our conclusion is that these alternative variables do not perform any better, and the $[M,1+\dsfof]$ pair shows the least systematic variation with redshift.

In summary, the simple parametrisation of the environmental dependence of the merger rate given by equations~(\ref{eqn:dNdz}) and (\ref{eqn:FIT}) can be used along with the fit for the global merger rate $B/n$ in equation~(\ref{eqn:Bnfit}) to compute the merger rate in different environments at a variety of redshifts.

\subsection{Progenitor Mass Function} \label{CMF}

For completeness and ease of comparison with analytic models, we present here the results for the environmental dependence of the conditional (or progenitor) mass function $\phi(M_1,z_1|M_0,z_0)$. This function gives the mean distribution of the progenitor masses $M_i$ at redshift $z_1$ for a descendant halo of mass $M_0$ at redshift $z_0$. It is the key ingredient for the construction of Monte Carlo merger trees in the Extended Press-Schechter model.

The relation between $\phi$ and the merger rate $B/n$ is discussed in Sec~3.3 of FM08. These two quantities are closely related but differ in two ways. First, $\phi$ is typically plotted vs $M_1/M_0$, while $B/n$ is expressed in the mass ratio of the progenitors $M_i/M_1 (i\ge 2)$ and the descendant mass $M_0$. Second, the conditional mass function $\phi(M_1,z_1|M_0,z_0)$ includes all progenitor halos at $z_1$ regardless of if a merger has occurred between $z_1$ and $z_0$, whereas the merger rate includes only descendant haloes with more than one progenitor. When the lookback time $z_1-z_0$ is small, a large fraction of haloes in fact have only one resolved progenitor typically with a mass $M_1$ comparable to the descendant mass $M_0$. See the sharp rise in $\phi$ near $M_1/M_0=1$ in Fig.~\ref{fig:CMF}). No such peak is present in the merger rate in Fig.~\ref{fig:Bn}.

Fig.~\ref{fig:CMF} shows that the progenitor mass function has a similar dependence on $1 + \dsfof$ as the merger rate in Figs.~\ref{fig:Bn}-\ref{fig:Bn2}. We have chosen to plot Fig.~\ref{fig:CMF} in the same way as Fig.~\ref{fig:Bn}, where the upper left panel shows the global progenitor mass function $\phi$ at $z_1=0.06$ for five bins of descendant mass $M_0$ at $z_0=0$, and the other five panels show how $\phi$ for haloes in different $\dsfof$ bins compare to the global mean $\phi$. We see that, like $B/n$, the progenitor mass function has a noted dependence on environment. For galaxy-size descendant haloes, those in the overdense regions have $\sim 1.5$ times as many progenitor haloes as the mean, while those in the underdense regions have $\sim 0.7$ times as many progenitor haloes as the mean.

\section{Alternative Algorithms for Post-Processing Halo Fragmentations} \label{Split}
\begin{figure}
	\centering 
	\includegraphics{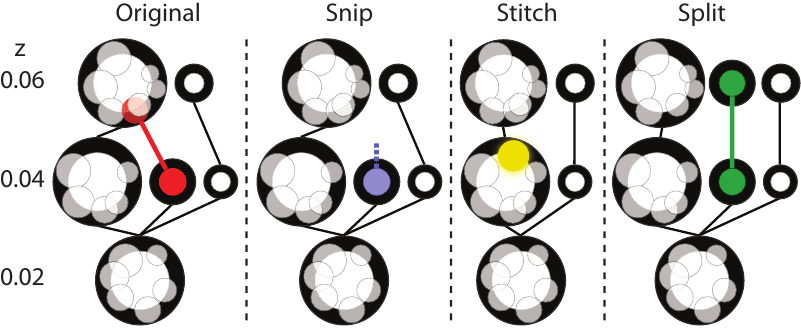} \caption{Example of a typical fragmentation event in the Millennium simulation. Black circles represent FOF haloes; white circles represent subhaloes. Circle radii scale with the logarithm of the (sub)halo mass. The left panel shows a fragmentation event occurring between $z=0.06$ and 0.04 (red subhalo). The snip panel shows how the ancestral link between the fragmented halo and its progenitor is severed, producing a (blue) orphan halo. The stitch panel shows how the fragment is stitched back into the main branch at $z=0.04$ (yellow subhalo). The split panel shows how the fragment's progenitor at $z=0.06$ is split off from the FOF halo (green subhaloes).} \label{fig:MergerTreeCartoon} 
\end{figure}

As we discussed in Sec.~2 and FM08, even though each subhalo in the Millennium tree is, by construction, identified with a single descendant subhalo, the resulting FOF tree can contain fragmentation events in which an FOF halo is split into two (or more) descendant FOF haloes. This fragmentation issue is not unique to the use of subhaloes in the Millennium simulation but occurs in merger trees in all prior studies that are typically constructed based on the FOF haloes rather than subhaloes. This problem arises because particles in a progenitor halo (or subhalo) rarely end up in exactly one descendant halo; a decision must therefore be made to select a unique descendant and there is no unique way to do this. The standard procedure to assign progenitor and descendant FOF haloes is the same as that applied to the {\it subhaloes} in Millennium: the descendant halo is the halo that inherits the most number of bound particles of the progenitor. We call this algorithm {\it snipping} since it effectively cuts off the ancestral link between a progenitor halo and its {\it subdominant} descendant fragments, while leaving the halo masses unchanged (see Fig.~\ref{fig:MergerTreeCartoon}).
\begin{figure*}
	\centering 
	\includegraphics{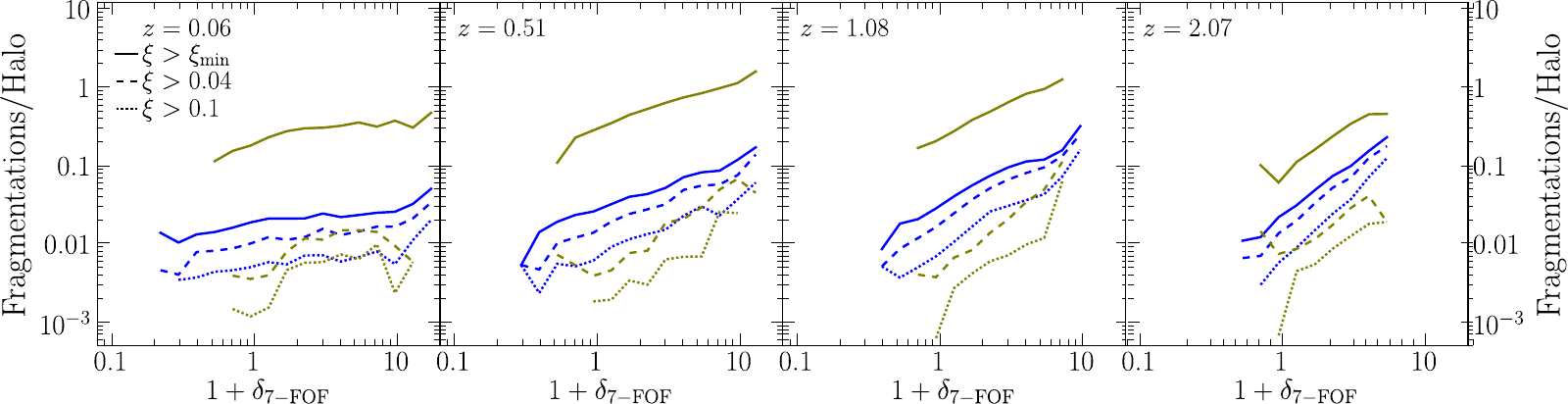} \caption{Fragmentation statistics (fraction of haloes that fragmented) as a function of environmental density $1+\dsfof$ at redshifts $z=0.06,\ZB,\ZC,\ZD$ (left to right). Within each panel, three fragmentation mass ratios $\xi$ are shown: major fragmentations with $\xi > 0.1$ (dotted); those with $\xi > 0.04$ (dashed), and all fragmentations down to the resolution limit (40 particles; solid). The two colours are for different mass bins: 0 to 40\% (blue) and 90-99\% mass bin (red); see Table~\ref{table:MassBins}. The red solid curve is significantly higher than the blue solid curve because the fragments of higher mass haloes are better resolved (i.e. $\xi_{min}$ is smaller). Minor fragmentations are seen to dominate, while only $\sim 1$\% of haloes suffer major fragmentations ($\xi > 0.1$) in typical environments.} \label{fig:FragStats} 
\end{figure*}

In FM08, we explored a new method \emph{stitching} for handling these fragmentation events. In this method (which we call {\it stitch-3} here), the fragmented haloes that remerge within 3 outputs after fragmentation occurs are stitched into a single FOF descendant; those that do not remerge within 3 outputs are snipped and become orphan haloes. We compared the two methods and showed that snipping inflates the merger rates by up to 10\% in the major merger regime and 25\% in the minor merger regime (Fig.~9 of FM08). This is not surprising since bound subhaloes are often on eccentric orbits that extend out to 2 to 3 virial radii of the main halo (see, e.g., \citealt{ludlow08}). The FOF finder can repeatedly disassociate and associate these subhaloes, leading to spurious fragmentation and remerger events.

In addition to the snipping and stitching algorithms, we examine a third method here that is complementary to stitching. We call this method {\it splitting} (see also \cite{Genel}). Our motivation for introducing this algorithm is the fact that fragmentations can be the result of either {\it false fragmentation} at the lower $z_0$, where physically bound subhaloes are broken up, or {\it false grouping} at the earlier $z_1$, where physically unbound subhaloes are falsely associated by the FOF finder. Even though multi-body subhalo encounters may unbind a subhalo, our visual inspections of a number of halo merger tracks indicate that such events are rare. Instead, most of the apparent fragmentations are due to the halo finder, which at an earlier output ($z_1$) may group subhaloes together, only to separate them at the next timestep ($z_0<z_1$). A decision needs to be made about whether the falsely separated haloes at $z_0$ should be put back together (i.e. {\it stitching}), or the falsely grouped halo at $z_1$ should be broken up (i.e. {\it splitting}). (Note: {\it Snipping} effectively does nothing.)
\begin{figure*}
	\centering 
	\includegraphics{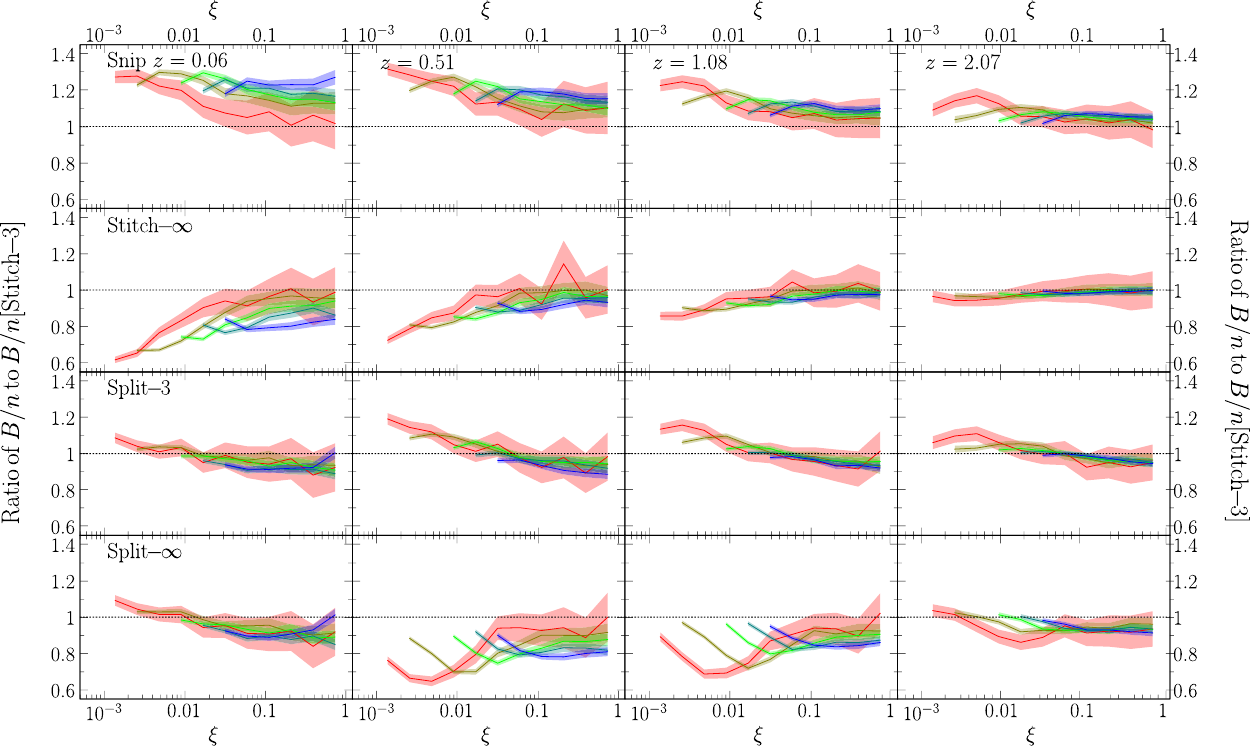} \caption{Comparison of the global merger rate $B/n$ (including all environment) vs progenitor mass ratio $\xi$ computed from five fragmentation algorithms at $z=0.06,\ZB,\ZC,$ and $\ZD$ (left to right). Since stitch-3 is the method used in FM08, we plot the ratio of $B/n$ from the other four methods to $B/n$ from stitch-3. Colours correspond to the mass bins in Table~\ref{table:MassBins} (blue: lowest mass, red: highest mass). The shaded regions denote Poisson errors.} \label{fig:BN_xi_alg} 
\end{figure*}
\begin{figure*}
	\centering 
	\includegraphics{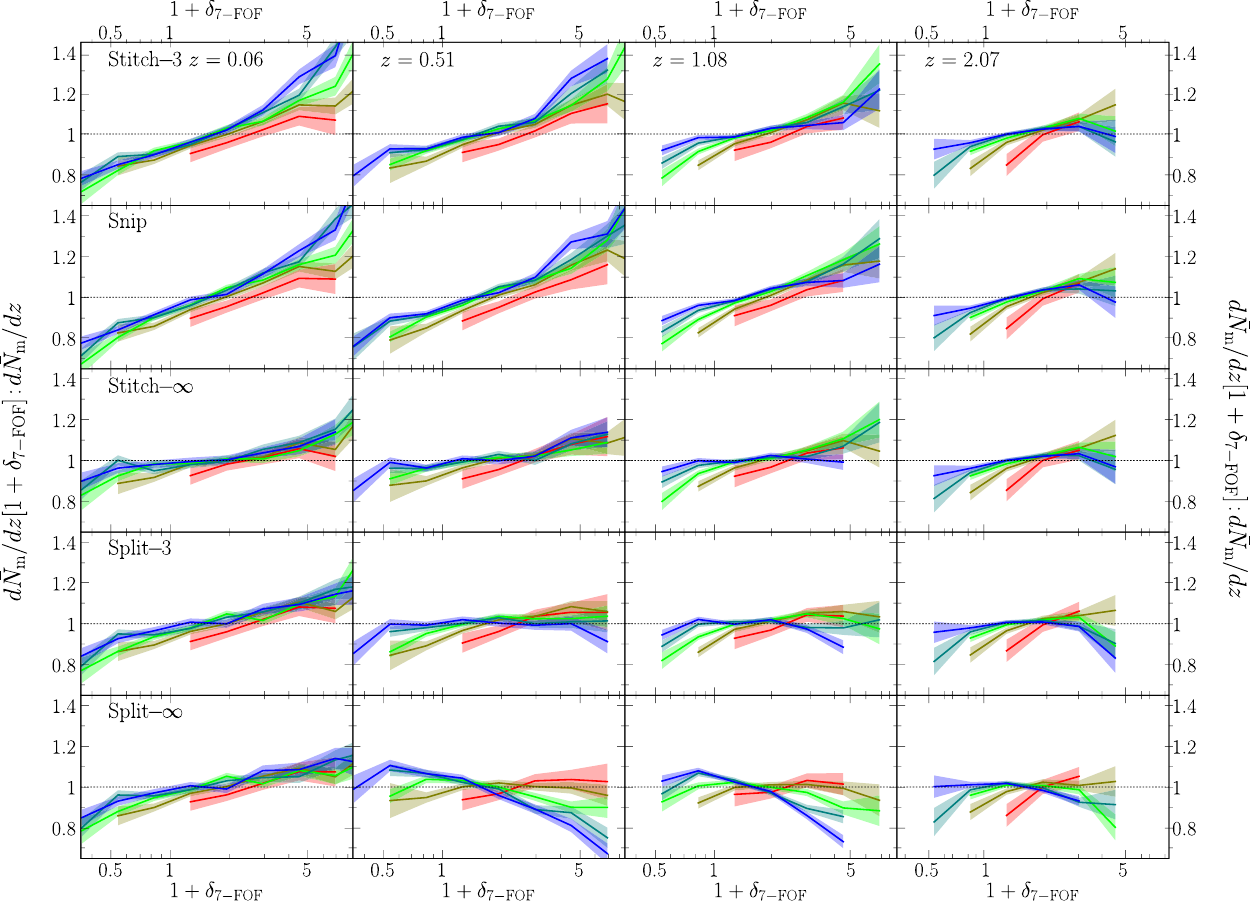} \caption{ Comparison of the environmental dependence of the merger rate computed from five algorithms at $z=0.06,\ZB,\ZC,\ZD$ (left to right). Similar to Fig.~\ref{fig:Bn2}, we plot the ratio of the mean merger rate $\Rz[\delta]$ for haloes in a given $\dsfof$ bin to the global mean merger rate $\Rz$. The mass bins are shown in different colours (blue for the lowest and red for the highest bin in Table~\ref{table:MassBins}). Shaded regions indicate Poisson errors. This figure shows that stitch-3, snip, and split-3 have similar $\delta$ dependence. Split-$\infty$, however, reverses the $\delta$-trend for low mass haloes, which we believe is an artefact of the propagation of fragmentations up the tree (see text).} \label{fig:BN_delta_alg} 
\end{figure*}

Fig.~\ref{fig:MergerTreeCartoon} illustrates how each of the three algorithms -- snip, stitch, and split -- handles halo fragmentation. For completeness, we also explore a variation of stitch (and split), in which the number of outputs used to make the decision is altered. Instead of stitch-3 (or split-3), which only stitches (or splits) fragmented haloes that remerge within 3 time outputs, we consider stitch-$\infty$ (or split-$\infty$), which stitches (or splits) any fragmented haloes regardless of their future (or past) history. An important distinction between stitch-3 and stitch-$\infty$ (and similarly for split-3 vs split-$\infty$) is that the modifications to the haloes are confined to the three adjacent outputs in stitch-3 and split-3; haloes along the merger tree outside of this time range are unaltered. The modifications made in stitch-$\infty$ and split-$\infty$ however, propagate indefinitely either forward or backward along any tree branch where a fragmentation occurs. No algorithm is perfect, but any error made in stitch-$\infty$ and split-$\infty$ will affect the entire branch of the tree that contains a fragmentation event. By contrast, errors made in stitch-3 and split-3 are confined to the redshift at which the fragmentation occurs. Stitch-$\infty$ and split-$\infty$ are therefore extreme algorithms, which are included here for comparison purposes only.

Before comparing the algorithms, we first show the frequency of fragmentations in the Millennium FOF tree as a function of environment in Fig.~\ref{fig:FragStats}. The fraction of haloes that experience fragmentations is seen to increase with $\dsfof$, differing by a factor of $\sim 3$ at low $z$ and by a factor of $\sim 5$ to 10 at $z\approx 2$. The fragments, however, are dominated by low-mass haloes: only $\sim 1$\% of the haloes in typical densities have fragments of mass ratio $\xi$ above 0.1, and this fraction is no larger than $\sim 10$\% even in the densest regions. Most of the fragmentations are therefore minor.

Fig.~\ref{fig:BN_xi_alg} compares the global mean merger rates $B/n$ (i.e. including all environment) as a function of progenitor mass ratio $\xi$ for the five algorithms (top to bottom) at four redshifts ($z=0.06, \ZB, \ZC, \ZD$ from left to right). Since stitch-3 is the algorithm used in FM08, we show the ratio of each of the four alternative algorithms to stitch-3. Within each panel, the coloured curves show a variety of descendant mass bins (the bands show Poisson errors). As we have already seen in FM08, snipping (first row in Fig.~\ref{fig:BN_xi_alg}) yields a higher merger rate (by $\sim 20$\% at $\xi < 0.01$) due to the orphaned haloes, resulting in a steeper power-law dependence $\xi^{\beta}$ ($\beta\sim-2.2$) than stitch-3 ($\beta\sim-2$). Stitch-$\infty$ (second row), on the other hand, zips together all the fragments and reduces the number of minor mergers by as much as $\sim 40$\% ($\beta\sim-1.8$) in comparison to stitch-3. Split-3 (third row) tends to raise the minor merger rate by up to $\sim 15-20$\%. Split-$\infty$ (fourth row) has a feature in the low-$\xi$ merger rate that breaks the power-law behaviour seen in the other trees. This feature is redshift dependent and drives the merger rate lower than in stitch-3.

We now examine how the environmental dependence of the merger rates is affected by the algorithm used for handling fragmentations. To do this, we integrate $B/n$ over $\xi$ and show the total rate, $\Rz$, as a function of $1+\dsfof$ for the five algorithms (top to bottom) at four redshifts in Fig.~\ref{fig:BN_delta_alg}. Similar to Fig.~\ref{fig:Bn2}, the vertical axis shows the ratio of the merger rate in a $\dsfof$ bin to the global rate, $\Rz[\delta]:\Rz$, computed with each method. This figure shows that the stitching (both stitch-3 and stitch-$\infty$) and snipping algorithms produce very similar environmental dependence; though the extreme stitch-$\infty$ yields a mildly weaker $\delta$ dependence.  In contrast, split-$\infty$ shows a sudden \emph{reversal} in the $\delta$-dependence at $z=\ZB, \ZC, \ZD$ in the three lower mass bins, with haloes in the densest regions experiencing {\it fewer} mergers.  Split-3, on the other hand, shows a positive (albeit weak) correlation of merger rate with $\delta$ for all mass bins but the very lowest. We believe the difference between the split and stitch trees is due to an "unzipping" effect that is most pronounced in split-$\infty$, in which splitting a fragmentation event at low $z$ affects the entire branch above this redshift, resulting in the discrepantly low merger rates in high density regions seen in the last row of Fig.~\ref{fig:BN_delta_alg}.

In summary, halo fragmentation is a generic feature of all merger trees. It occurs more frequently in dense regions than in voids, thereby prompting the detailed investigation in this section. Our tests of five algorithms show that the majority of tree-processing methods (stitch-3, stitch-$\infty$, snip, and, to some extent, split-3) give very similar environmental dependence for the mean merger rate. In addition, the global merger rate (including all environment) is robust, differing by less than 10\% for major mergers and less than 20\% even in the very minor merger regime ($\xi < 0.01$) that is more prone to systematic effects. The split-$\infty$ algorithm, on the other hand, appears to suffer from ``non-local'' effects that have propagated up the merger tree from the fragmentation point. In particular, the merger rate is greatly reduced in high density regions when split-$\infty$ is used.

\section{Conclusions and Implications} \label{Conclusions}

We have used the dark matter haloes and merger trees constructed from the Millennium simulation to quantify the dependence of halo merger rates on halo environment from redshift $z=0$ to 2. A number of local mass density parameters centred at the haloes, both including and excluding the central halo mass itself, are tested as measures of environment. We have found that $\dsfof$ defined in equation~(\ref{delta3}) is a robust measure of the surrounding environment outside of a halo's virial radius. It cleanly subtracts out the contributions to the local density from the central halo and thereby breaks the degeneracy between halo mass and environment for high mass haloes (see Figs.~\ref{fig:DeltaComparison} and \ref{fig:DeltaDistribution}).

We have found strong and positive correlations in both the halo merger rate and the progenitor mass function with environmental densities. Figs.~\ref{fig:Bn}-\ref{fig:CMF} present our main results, where haloes in the densest regions are seen to experience 2 to 2.5 times higher merger rates than haloes in the voids. Such a density dependence can be approximated analytically by multiplying our earlier fitting formula FM08 for the global merger rates (eq.~\ref{eqn:Bnfit}) by an additional $\delta$-dependent factor given by equation~(\ref{eqn:FIT}). This factor is a simple power-law in both the environmental density and halo mass, and it is redshift-independent. The mass dependence is quite weak, indicating that haloes with different masses but similar values of $1+\dsfof$ experience similar merger rate amplifications. This is intriguing in light of the fact, discussed in Section~\ref{Detangling}, that these haloes actually reside in different environments.

The strong correlations of the halo merger rate and progenitor mass function with environment discussed in this paper have important implications for the analytic Press-Schechter \citep{PS74} and excursion set models \citep{BondEPS, LC93}. In this popular formalism, halo growth is modelled by the random walk trajectories of dark matter density perturbations smoothed at decreasing scales. Haloes are identified at scales at which these trajectories first cross some critical density threshold, and the Markovian nature of the model allows one to compute the distribution of these first crossings. This distribution is then mapped onto the number-weighted conditional mass function $\phi(M,z|M_0,z_0)$ discussed in Section~\ref{CMF} and plays an important role in the Monte Carlo construction of mock merger tree catalogues (see \citealt{ZFM08} and references therein).

It is generally assumed that the conditional mass function is independent of environment as the excursion set model is Markovian. The Markovian nature of the random walks, however, is not a prediction but rather an assumption resulting from the use of the $k$-space tophat window function to smooth the density perturbations. There have been recent attempts to weaken this assumption or to introduce environmental dependence into other parts of the model \citep{ZentnerEPS, Sandvik07, DesJacques07}, but these modifications thus far have not been able to reproduce the basic statistical correlation between halo clustering and formation time found in simulation studies: older haloes are more clustered \citep{Gottlober01, ShethTormen04, Gao2005, Harker06, Wechsler06, JingSutoMo07, WangMoJing07, GaoWhite07, Maulbetsch07}.

How do our environmental results for the merger rates tie in with these simulation and EPS studies? We have shown that the amplification of halo merger rates in denser regions persists at all redshifts (up to at least $z=2$). If mergers were the dominant channel for halo growth, our results would imply that for haloes of a {\it fixed} mass today, those in denser regions should have formed {\it more recently} than those in void regions. Interestingly, this is exactly opposite to the trend reported in many recent studies that have found older (i.e. earlier forming) haloes to be more clustered than younger haloes. As we will discuss in the next paper (Fakhouri \& Ma 2008c), these two results are in fact not in conflict once the other important channel for halo mass growth -- the ``diffuse'' accretion of non-halo material (either unresolved or stripped) -- is taken into account. We will quantify the environmental dependence of this component and show that, when combined with the merger rate results presented in this paper, we recover the formation redshift dependence reported in prior simulation studies.

\section*{Acknowledgements}

We thank Simon White, Jun Zhang, and the referee for useful comments. This work is supported in part by NSF grant AST 0407351. The Millennium Simulation databases used in this paper and the web application providing online access to them were constructed as part of the activities of the German Astrophysical Virtual Observatory.

\bibliographystyle{mn2e} 
\bibliography{env1}
\begin{appendix}
	\section{Self-Similar Mass and Environment Variables} \label{AppendixSelfSim}
	\begin{figure}
		\centering 
		\includegraphics{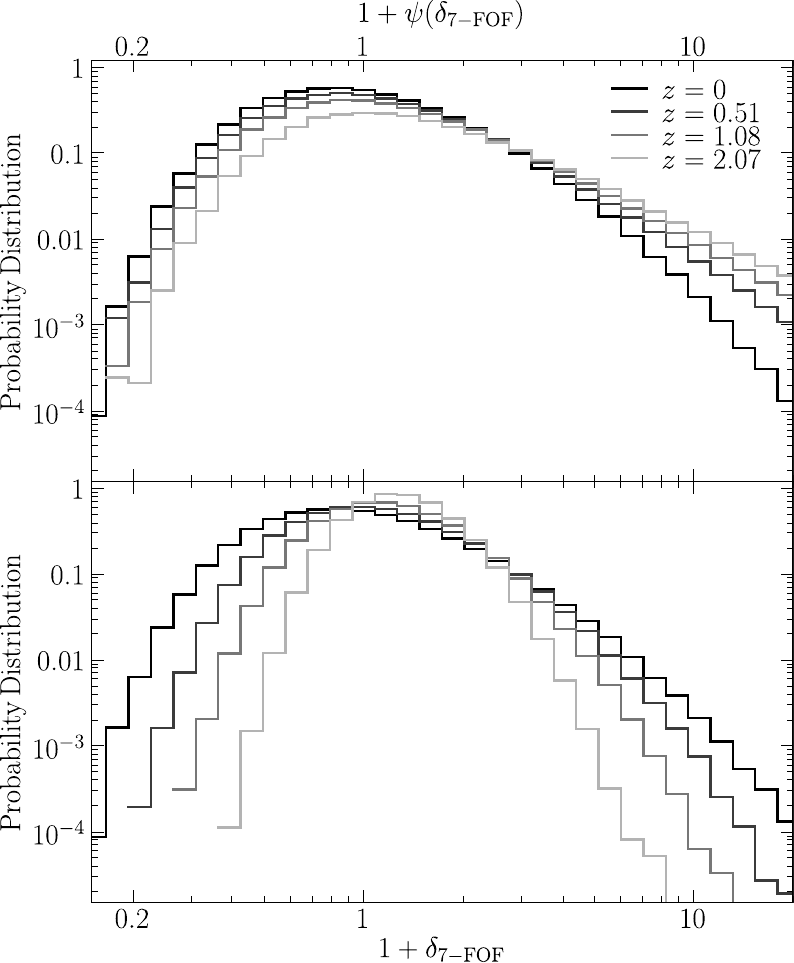} \caption{Distribution of scaled overdensity variable $1+\psi(\delta_{7-\rm{FOF}})$ (top panel) vs the original $1+\dsfof$ (bottom panel) for $z=\ZA,\ZB,\ZC,$ and $\ZD$. Each histogram is normalised to have unit area. The broadening of the $1+\dsfof$ distribution with decreasing $z$ is a natural consequence of gravitational instability. This evolution is largely removed when $\psi(\delta_{7-\rm{FOF}})$ is used as the variable, suggesting that $\psi(\delta_{7-\rm{FOF}})$ is a more self-similar measure of halo environment.} \label{fig:AppendixDelta} 
	\end{figure}
	
	\begin{figure*}
		\centering 
		\includegraphics{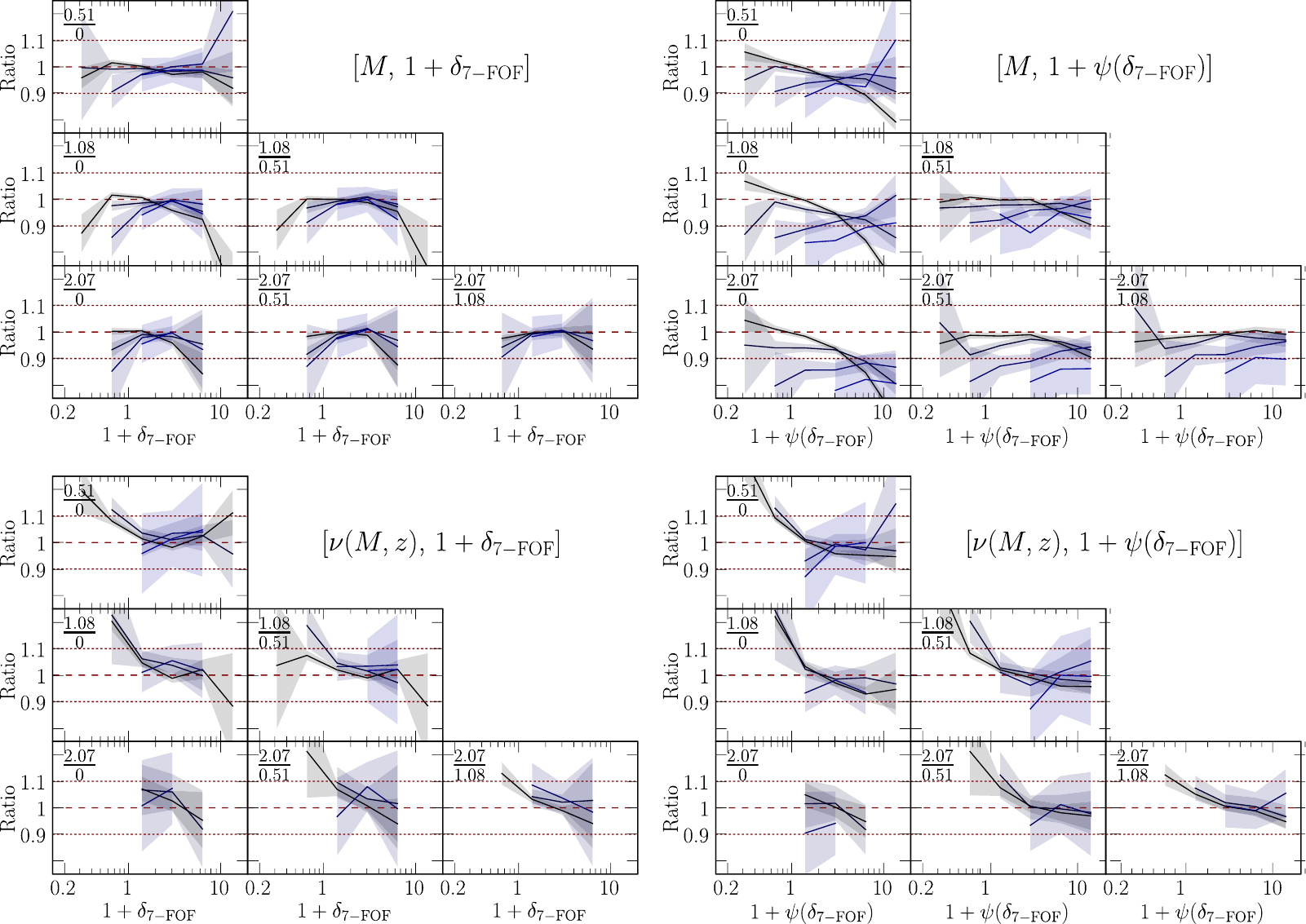} \caption{Ratios of the environmental dependence of the halo merger rates, $R(M,\delta,z)=\Rzf[M,\delta,z]/\Rzf[M,z]$, computed at different redshifts: $R(z_2)/R(z_1)$. Two mass and two environment variables are used, in clockwise order from the top-left matrix of subplots: ($M,\dsfof $),($M,1+\psi(\dsfof)$),($\nu(M,z),1+\psi(\dsfof)$), ($\nu(M,z),\dsfof$). Each matrix presents ratios of $R(z_2)/R(z_1)$ plotted as a function of environment. Each line represents a different mass bin with low mass in black and high mass in blue. The shaded regions represent Poisson errors. The redshifts used to compute $R(z_2)/R(z_1)$ are noted in the top left corner of each subplot. The top-left matrix ($M,\dsfof$), the variables used throughout this paper, shows the least systematic dependence on mass and environment.} \label{fig:AppendixRatio} 
	\end{figure*}
	
	We have chosen to use the intuitive mass and environment variables $M$ and $\dsfof$ in the fitting form (eq.~\ref{eqn:FIT}) for the environmental dependence of the merger rate for $0\leq z \leq 2$. Here we investigate if other choices of mass and density variables may improve the fit. This is motivated by the well known property that the unconditional halo mass function is (almost) redshift-independent when the variable 
	\begin{equation}
		\nu(M,z)=\frac{\delta_c}{\sigma(M) D(z)} 
	\end{equation}
	is used to characterise mass (see, for example, \citealt{Jenkins}); whereas when $M$ is used as the mass variable, the halo mass function evolves significantly with redshift. Here $\delta_c=1.686$ is the critical overdensity for spherical collapse, $\sigma(M)$ is the variance of the linear density perturbations evaluated at a scale corresponding to the halo mass $M$, and $D(z)$ is the linear growth function. On the other hand, as we discussed in the paper, the merger rate is more closely related to the {\it conditional} mass function than the {\it unconditional} mass function, and the redshift dependence of the former cannot be scaled out simply by using $\nu$. Nonetheless, one can ask whether $\nu$ is the more appropriate variable for capturing the mass dependence of the environmental dependence of $\Rz$.
	
	A similar question can be raised about $\dsfof$. The overdensity $\dsfof$ grows as a result of gravitational instability, leading to broader distributions of $\dsfof$ towards lower redshifts as shown in Fig.~\ref{fig:DeltaDistribution}\footnote {We have also compared distributions across fixed $\nu$ bins and found equivalent changes in the distribution of $\dsfof$.}. We can scale out the growth of $\dsfof$ in the linear regime by replacing $\dsfof(z)$ by $\dsfof(z)/D(z)$.
	
	To incorporate the effects of nonlinear growth, we use the approach of \cite{GV04} for underdensities and \cite{PEEBLES84} and \cite{EKE96} for overdensities. These authors assume that the under/overdense regions are spherically symmetric and apply Birkhoff's theorem, treating these regions as self-contained universes embedded within an expanding universe. The model cosmological parameters for these embedded cosmologies are computed from the density of the region under consideration, and the resulting Friedmann equation is solved to relate the densities at some redshift, $\dsfof(z)$, to densities, $\psi(\dsfof(z))$, at $z=0$. When $\dsfof\ll1$, this procedure is in agreement with the linear relation $\psi(\dsfof)\sim \dsfof/D(z)$
	
	A further complication for $\dsfof$ is that the mass of the central object has been removed from $\delta_7$. We have tested swapping the order of operation by first nonlinearly propagating $\delta_7\rightarrow \psi(\delta_7)$ {\it then} subtracting the mass of the central object. We find that the resulting distributions of $\psi(\delta_7)_{-\rm{FOF}}$ are only slightly modified from the distributions of $\psi(\dsfof)$.
	
	Fig.~\ref{fig:AppendixDelta} compares the distribution of the scaled $1+\psi(\dsfof)$ (top panel) with that of the original $1+\dsfof$ (bottom panel) for all haloes with $M>\MMIN M_\odot$ at $z=\ZA,\ZB,\ZC$ and $\ZD$ (black to light grey lines). The upper panel shows far less broadening with decreasing $z$ than in the lower panel, indicating that $\psi(\dsfof)$ {\it does} remove much of the redshift evolution of $\dsfof$. The mapping is clearly imperfect: The distributions at $z=\ZC$ and $z=\ZD$ have long positive density tails that are not present at $z=0$. This is not surprising since the simple spherical approximation used for evolving the density cannot account for all non-linear effects such as the mergers of overdense and underdense regions.
	
	To test if $\psi(\dsfof)$ and $\nu$ are more appropriate variables to use in the fitting formula than $\dsfof$ and $M$, we show in Fig.~\ref{fig:AppendixRatio} the ratio of $\Rz[\delta]$ to the global mean $\Rz$ for a number of log-spaced mass bins at $z=\ZA,\ZB,\ZC,$ and $\ZD$. For brevity, let us refer to this ratio as $\RATIO(M,\delta,z)$. We can compute $\RATIO$ using either $M$ or $\nu(M,z)$ as the mass variable, and either $\dsfof$ or $1+\psi(\dsfof)$ as the environmental variable. The upper left set of plots in Fig.~\ref{fig:AppendixRatio} uses $M$ and $1+\dsfof$, which are the variables used throughout this paper; the upper right set uses $M$ and $1+\psi(\dsfof)$; the lower left set uses $\nu(M,z)$ and $1+\dsfof$; and the lower right set uses $\nu(M,z)$ and $1+\psi(\dsfof)$.
	
	Within each set of figures we plot a matrix of ratios of $\RATIO$ computed at different redshifts. The redshifts used to compute the ratios are noted in the upper left corner of each subplot. For example, the upper subplot is the ratio $\RATIO(M,\delta,\ZB)/\RATIO(M,\delta,\ZA)$ and is labelled ``$\ZB/\ZA$''. Each subplot contains five mass bins. Only points containing more than 40 haloes are plotted to minimise noise (this results in some mass bins being dropped). The variables that successfully capture the redshift evolution in the merger rate will show very little variation in $\RATIO(M,\delta,z)$ with redshift and give ratios $\RATIO(M,\delta,z_1)/\RATIO(M,\delta,z_2)$ close to unity. Interestingly, the $[M,\dsfof]$ pair used throughout this paper does the best job. The upper left matrix in Fig.~\ref{fig:AppendixRatio} shows ratios of $\RATIO$ that tend to cluster around $1$ and show few systematic trends with mass and environment.

	Using $\nu(M,z)$ instead of $M$ (lower panels) introduces a strong $\delta$ dependence in the ratios of $\RATIO$: the $\delta$-slope of $\Rz$ flattens with increasing redshift. Similarly, using $1+\psi(\dsfof)$ instead of $\dsfof$ (right panels) introduces a strong mass dependence when $M$ is used as mass variable and does not improve on the $\delta$ dependence introduced by $\nu(M,z)$.
	
	Thus, $M$ and $1+\dsfof$ appear to be the optimal variables for capturing the environmental dependence of dark matter halo merger rates for $0\leq z\leq2$.
\end{appendix}

\label{lastpage}

\end{document}